\documentstyle[psfig]{mn}
\title[Clustering, Counts and Morphology of Extremely Red Galaxies]
{The Clustering, Number Counts and Morphology of Extremely Red
($\bf R-K>5$) Galaxies to $\bf K\leq 21$} 
\author[N.D. Roche, O. Almaini, J.S. Dunlop, R.J. Ivison,
C.J. Willott]
{Nathan D.
Roche$^{1,4}$, Omar Almaini$^{1,5}$, James Dunlop$^{1,6}$,
 R.J. Ivison$^{2,7}$ \and and C.J. Willott$^{3,8}$\\
$^1$Institute for Astronomy,
     University of Edinburgh,
     Royal Observatory, 
     Edinburgh EH9 3HJ,
     Scotland.\\
$^2$ Astronomy Technology Centre,
     Royal Observatory, 
     Edinburgh EH9 3HJ,
     Scotland.\\
$3$  Astrophysics,
     Department of Physics,
     Keble Road,
     Oxford OX1 3RH,
     England.\\
{$^4$ \verb"ndr@roe.ac.uk"}\hspace{8mm}   
{$^5$ \verb"omar@roe.ac.uk"}\hspace{8mm}
{$^6$ \verb"jsd@roe.ac.uk"}\hspace{8mm}
{$^7$ \verb"rji@roe.ac.uk"}\hspace{8mm}
{$^8$ \verb"cjw@astro.ox.ac.uk"}\hspace{8mm} 
}

\begin{document}

\maketitle
\begin{abstract}
Using $K$ and $R$ band imaging of the ELAIS N2 field, 
we investigate the number counts, clustering, morphology and
radio/X-ray emission of extremely red objects (EROs), defined as
galaxies with $R-K>5.0$. This criterion will select old, passive
ellipticals at $z>0.9$. To $K=21$ we identify a total of  
158 EROs in 81.5 $\rm arcmin^2$. The ERO  number counts are lower than
predicted by pure luminosity evolution models, but  higher than
predicted by current CDM-based hierarchical models. The ERO counts
are
 consistent
with a non-evolving model and also 
with a luminosity evolution model incorporating 
moderate merging and a decrease with redshift in the comoving
number density of passive galaxies (`M-DE'). 

We investigate the clustering of the EROs by calculating their 
angular correlation
function, $\omega(\theta)$, and  obtain a $>2\sigma$ detection of
clustering at $K=19$--20 limits. The $\omega(\theta)$ amplitude of
these  EROs is 
 much higher than that of  
full $K$-limited samples
of galaxies, and best-fitted by models with a   comoving
correlation radius 
$r_0\simeq 10$--$13~h^{-1}$ Mpc.
 These results, which are in agreement with Daddi
et al. (2000),
 suggest that the intrinsic clustering of at least the brighter EROs
is even stronger than
that of present-day giant ellipticals.

We estimate seeing-corrected angular sizes and morphological types
for a $K\leq
19.5$ subsample of EROs (31 galaxies) and 
find a $\sim$3:2 mixture of
bulge and disk profiles. Of these EROs $\sim {1\over 4}$  appear to be
 interacting, disturbed or otherwise
irregular, and two are visible  mergers.
 We find the angular sizes of the bulge-profile EROs are consistent with 
passively evolving ellipticals in the M-DE model, at the expected
 $z\sim 0.9$--2. The ERO mean radii  are smaller than the non-evolving
prediction, implying surface brightness evolution.
 
Seven of the 31 bright EROs are detected  as 
$F(1.4~{\rm GHz})\geq 30\rm \mu Jy$ radio sources in a VLA survey.
 The strongest, at 5 mJy, is also
 a Chandra X-ray detection, and lies at the centre of 
a significant overdensity of EROs -- it is probably an FRI radio galaxy
in a $z\sim 1$ cluster. Of the other, much fainter,
sources, five  are point-like
and may be weak AGN, while the sixth is elongated and aligned with the
optical axis of an  extended,
 low-surface brightness ERO, and hence probably a ULIRG-type
starburst.

A possible interpretation is discussed in which the EROs are a mixture
of (i) `pEROs', strongly clustered passively evolving giant ellipticals,
 formed at high redshifts, the oldest EROs, and (ii) `dsfEROs', dusty
  post-interaction
galaxies, with a few active starbursts (ULIRGs), and less strongly
clustered.
With time,    
the younger dsfEROs are continually assimilated into to the ERO class, diluting
the clustering and increasing the comoving number density. Both
types ultimately evolve into  today's early-type galaxies.
\end{abstract}

\section{Introduction}
Deep  surveys at  near-infra-red  ($\sim 2 \rm \mu  m$) wavelengths  
revealed that  a population  of very red  ($R-K>5$ or
$I-K>4$, Vega system) galaxies appears faintward of $K\sim
18$ (e.g. Elston et al. 1988; Hu and Ridgeway 1994). The colours and
$K$ magnitude range of these `extremely red objects'
(EROs) are consistent with old, passive
 (i.e. no longer forming new stars) galaxies, observed 
at $z\sim 1$ or beyond.
These would be the progenitors of
present-day giant ellipticals, 
thought to have formed  their stars  at even
higher ($z\geq 3$) redshifts.

The EROs are of great cosmological
interest, as  they will include  the earliest formed of  all galaxies,
and their properties will help to answer the long-standing mysteries
concerning the early evolution of elliptical galaxies (e.g. Jimenez et
al. 1999). One hypothesis is that the present-day
ellipticals formed in  single starbursts  at $z>3$,
thereafter undergoing  only passive luminosity evolution, as in pure
luminosity evolution (PLE) models, another is that they formed more recently
through
mergers of spiral galaxies, or by some combination of the two. 
Initial surveys  of EROs  (e.g. Barger et  al. 1999) found  fewer than
expected from PLE models, and implied that $\leq
50$ per cent  of the local E/S0s could have formed  in a single high-redshift
starburst.  However, the detection  with SCUBA of many sub-mm sources, with
optical magnitudes
and  colours  consistent with  dusty  galaxies  at  $z>2$,
and  $850\mu \rm m$  fluxes indicative of  star-formation rates  (SFRs) as
high as $\rm \sim 1000 M_{\odot}yr^{-1}$, provided
strong evidence 
 that some massive galaxies did form in intense early starbursts
(e.g. Smail et al. 1999). The passive EROs could then be the intermediate stage
 between the SCUBA sources and  local giant ellipticals.

 With sufficient dust-reddening, high-redshift star-forming galaxies
 may also have 
 $R-K>5$.  Spectroscopy of a few of the brightest EROs revealed that
 both old,
 passive (e.g. Dunlop et al. 1996; Spinrad et al. 1997;  Stanford et al.  1997) and dusty
 starburst (e.g.  Dey et  al. 1999; Smith  et al. 2001)  galaxies were
 present, but the relative proportions remained uncertain.  

The morphologies of the EROs provide further information on their nature. 
Moriondo, Cimatti  and Daddi (2000)  fitted radial profiles to  WFPC2 and
NICMOS images  of 41  EROs (to $K\sim  21$) and estimated  that 50--80
percent  were elliptical-like and  15 percent irregular or interacting.
Stiavelli and  Treu (2000) studied 30  EROs to $H\sim  23$, using 
NICMOS, and classified these into E/S0, disk, irregular and point-like
morphologies in  a ratio 18:6:3:3.   They assumed that only  the first
type were  the progenitors of local  E/S0s, and by comparing the counts of
these with a
PLE model (with $\Omega_m=0.3$, $\Omega_\Lambda=0.7$), estimated 
only $\sim 15$ per
cent of the present-day number of E/S0s could have
formed at $z\geq 3$. 
 
Daddi  et   al.  (2000)  presented  the  first   measurements  of  ERO
clustering. They found the angular
correlation function, $\omega(\theta)$, of  $K\leq 19.2$ EROs  to
be almost an  order of magnitude higher than  that of all 
galaxies  to   the  same $K$ limit.  As  local   giant  ellipticals  are
intrinsically much  more clustered than  disk galaxies (e.g.  Guzzo et
al.  1997),  this  was interpreted  as  evidence  that  most  EROs  are
$z>1$ ellipticals.

Manucci et al. (2001) attempted a photometric separation of 
passive and dusty, star-forming EROs -- hereafter pEROs and dsfEROs --
on a plot of $J-K$ against $R-K$. Within a sample of  57 EROs with $K\leq
20$, they assigned equal numbers (21) to each class.
Cimatti et al. (2002a) provided the first true 
measurement of the dsfERO/pERO ratio, identifying by spectroscopy
approximately equal numbers (15 `dsf', 14 `p') amongst 45 EROs with $K\leq
19.2$. Firth et al. (2002) investigated ERO clustering to $H=20.5$
($K\simeq 19.5$) and obtained similar results to Daddi et al. (2000).
 
In this paper we select a sample  of $R-K>5.0$ EROs from new $K$ and $R$
band   images,   investigate  their   number   counts,  clustering   and
morphologies and compare these results with evolutionary models.
We do not  yet have spectroscopy for these galaxies,  but do have deep
radio  and  X-ray  observations  which  may  identify 
starbursts  and  AGN. 
Magnitudes  in this  paper are  given in  the Vega  system and  can be
converted   to  the   AB  system   using   $R_{AB}=R_{Vega}+0.19$  and
$K_{AB}=K_{Vega}+1.87$. Quantities dependent  on Hubble's Constant are
given in terms of $h_{50}=H_0/50$ or $h_{100}=H_0/100$ km $\rm s^{-1}Mpc^{-1}$.

This paper is organized as follows: Section  2  
describes  the dataset  and  its
reduction and analysis, and Section 3  the selection of the EROs and 
their distribution on the sky. In Sections
4  and  5 we  investigate,  in turn,  the  number  counts and  angular
correlation function,  $\omega(\theta)$, of  the $K\leq 21.0$ EROs and
other galaxies, and  compare with models.  In Section 6
we fit radial profiles to a subsample of  brighter, $K\leq 19.5$, EROs to
 investigate their morphologies and angular sizes, and in Section 7
describe the radio properties.
8  discusses  these  findings and with other recent
studies of EROs , and  their
possible interpretation.

\section{Observations}
\subsection{Data: {$\bf K$} and {$\bf R$} Imaging}
Our $K$ ($\lambda=2.2\rm \mu m$) and $R$ band imaging 
 covers an area within field N2 of the European Large
Area ISO Survey (ELAIS, Oliver et al. 2001), centered at
R.A. $16^h 36^m 30^s$ Dec +41:04:30.
The first part of the $K$-band data
consists of a mosaic of 16 contiguous fields observed using the 
UKIRT Fast-Track Imager (UFTI), on Mauna Kea, Hawaii, between
1 January and 31 May 2000. Most frames received a total of 
 8000 seconds exposure time. The
UFTI camera
 contains a 
$1024\times 1024$ pixel HgTeCd array which, with a pixel size of
0.091 arcsec, covers $1.55\times 1.55$ arcmin. 

The second part consists of three fields observed
with the Isaac Newton Group Red Imaging Device (Ingrid) on the William Herschel
 Telescope (WHT), La Palma. The Ingrid fields lie on the edges of the
UFTI  mosaic, overlapping it slightly.
Ingrid is fitted with a $1024\times 1024$
 pixel near-IR detector with a pixel size of 0.238 arcsec, covering
 $4.06\times 4.06$ arcmin.
Our Ingrid data are not quite as deep as the UFTI mosaic,  due to a shorter 
 exposure time  (5540 sec) and lower instrumental sensitivity.

The UFTI and  Ingrid observations lie within the  area of an $R$-band
image obtained with the Prime focus camera on the WHT (May 1999). This
instrument contains 2 mosaiced $2048\times 4096$ pixel EEV chips, with
a pixel size of 0.238  arcsec, covering  $16\times 16$
arcmin. The $R$-band data, obtained in  May 1999 as part of a study of
faint Chandra  X-ray sources  (Gonz\'{a}lez-Solares et al.  2002),
 consists  of 13
spatially dithered 600  sec exposures, combined  to give an
octagonal $16.4\times 16.9$ arcmin frame with detection limit $R\simeq
26$. This has accurate  astrometry (to
$\leq 0.5$ arcsec) based on the positions of radio sources.

\subsection{Radio/X-ray/sub-mm Data}
The ELAIS  N2 field has been surveyed in a number  of other passbands.
Radio  (VLA)  observations (Ivison  et  al.  2002)  reach a  $3\sigma$
detection  limit of $F(1.4~{\rm  GHz})=30\rm \mu  Jy$ (with  beam size
$1.40\times 1.46$  arcsec), and detect $>100$  sources.  The available
X-ray  data consist of  a 75ks  Chandra observation  made on  2 August
2000,  covering a  $16.9\times 16.9$  arcmin field, in  which 91
sources  were detected to  $F(0.5$--$8.0 keV)\simeq  6\times 10^{-16}$
ergs $\rm s^{-1}cm^{-2}$  (Manners et al. 2002; Almaini  et al. 2002),
most of which now have  optical identifications (Gonz\'{a}lez-Solares
 et al. 2002).
Sub-mm  observations   with  SCUBA,   detected  17  sources   above  a
$3.5\sigma$ limit $F(850\rm \mu m)\simeq 8$ mJy (Scott et al. 2002).
\subsection{Data Reduction}
The UFTI and Ingrid data were reduced and analysed using {\sevensize IRAF}.
The UFTI dataset was made up of 
 16 pointings, each observed for 9
 spatially dithered
exposures of 800 sec (two pointings had slightly different 
 exposures of 700 sec and 1100 sec and were renormalized to
800 sec).  Our  intention was to combine all
 $9\times 16=144$ exposures into a single mosaiced image, and
 to this end, the   spacing between the grid of
pointings had been matched to the field-of-view and dither pattern so
that the
resulting mosaic would have an approximately uniform 7200 sec coverage
(except at the edges).
 
The sky background was subtracted from all UFTI data (using
{\sevensize IRAF} `sky').
Astrometry was  derived for each pointing, by  reference to detections
on the  WHT $R$  image. However, even with these transforms, 
 the raw  exposures could  not be  mosaiced  with acceptable
accuracy,  indicating slight  rotations or
other distortions between them.  This problem was remedied by choosing
one pointing as a reference  and then using {\sevensize IRAF} `wcsmap'
to fit  a general polynomial transform between  its co-ordinate system
and that  of the other 15.  These 15 frames were  then rebinned, using
{\sevensize  IRAF}  `geotran',  into  the co-ordinate  system  of  the
reference.  The rebinned  data could  then be  simply  combined, using
{\sevensize  IRAF}  `combine'   (with  `offsets=wcs'),  into  a
16-frame UFTI  mosaic (taking care  to exclude any `bad'  regions from
each exposure in the summation). The  final step was to trim the noisy
edges  (which received  less than  the  full exposure  time) from  the
mosaic, to leave a total usable area of 38.7 $\rm arcmin^2$.

The three fields of Ingrid data overlap slightly with the UFTI mosaic,
but not with each other, and were reduced separately. The data for each field
consisted of 90 spatially dithered exposures of 61.56 seconds (each in
turn made up of 6 exposures which were co-added during observing). 
All exposures were
debiased, and flat-fielding was performed for blocks of 9 consecutive
exposures, by dividing them by their median image (derived 
using `combine' with sigclip 
rejection, and normalized to a mean of unity). 
 
The spatial dithers, which followed the same sequence for each block
of 9 exposures, were measured using `xregister'. All 90 exposures of
each of the three fields could then be added using
`combine'. Magnitude calibration was determined from
short exposures of UKIRT standard stars, interspersed with the
observations. We then matched $\sim 10$--15 stars on each Ingrid field 
(detected as described below) to their known
RA and Dec on th WHT $R$ image and fitted astrometric transforms using
{\sevensize IRAF} `pltsol' (with rms residuals 0.11--0.25 arcsec).
\subsection{Source Detection} 

Sources were detected on the $K$ and $R$ images using SExtractor  (Bertin and
 Arnauts 1996). Throughout, magnitudes given are the
 `total' magnitudes derived by
 SExtractor by fitting elliptical apertures to each individual  detection.
 
 For the UFTI mosaic, our detection criterion was
 that a source must exceed a $1.5\sigma_{sky}$ threshold (21.36 $K$
 mag $\rm arcsec^{-2}$) in 16 contiguous pixels (0.13 $\rm arcsec^2$),
 and we also used a
 detection 
filter of Gaussian FWHM 4 pixels. Using a plot of detection
 magnitude against FWHM, stars could be separated from galaxies to
 $K=16.5$ and many spurious noise detections could be rejected.  
 The mean FWHM of unsaturated stars was only 0.67 arcsec, indicating
 very good seeing, although there is significant variation in the 
PSF between the different frames of the mosaic.

For the WHT $R$-band image, our detection threshold was
 $1.5\sigma_{sky}$ (26.66 $R$ mag $\rm arcsec^{-2}$)
 in a minimum area 4 pixels (0.23 $\rm arcsec^2$),
 with a detection filter of 2.0 pixels
 Gaussian FWHM. The total usable area
 is 212 $\rm arcmin^2$. The mean stellar FWHM
 is 0.76 arcsec, and from the number count of detected galaxies,
 detection appears to be  complete to $R\simeq 25.5$, with moderate (25--30 per
 cent) incompleteness at $25.5<R<26$  and a  turnover at $R>26$.
The Ingrid data (total area 49.5 $\rm arcmin^2$) 
have a similar pixel size and again we used the detection criterion of
 $1.5\sigma_{sky}$ (20.60--20.65 $R$ mag $\rm arcsec^{-2}$)
 in $\geq 4$ pixels, with a Gaussian 2.0 pixel
  FWHM filter. 
Stars (45) were separated from galaxies to
 $K=16.5$. The mean stellar FWHM of 0.68 arcsec again indicated good
seeing. 
The galaxy counts from the Ingrid data are 
 very similar to the UFTI
counts  to $K=19.5$ but turn over at $K>20$ rather than at $K>21$, hence 
 are $\sim 1$ mag less deep.

The next step was to merge the Ingrid and UFTI catalogs, by
 flagging for exclusion the Ingrid detections
 within the area already covered by the UFTI
 mosaic. The Ingrid and UFTI magnitudes of the overlap-region objects
 were generally consistent within $\sim 0.1$ mag. 
Excluding overlaps reduced the total catalog area to 83.6 $\rm
 arcmin^2$. Hereafter we make use of the combined UFTI+Ingrid catalog
 to $K=20$ and the UFTI mosaic only at $20<K<21$.

 \section{Identifying the $\bf R-K>5$ Galaxies}
\subsection{The Significance of $\bf R-K>5$}
Galaxies with observer-frame $R-K>5.0$ ($R-K> 3.31$ in the
AB system) are either old, passive (zero or near-zero SFR)
galaxies at $z\geq 0.9$, or are very dust-reddened.
 Figure 1 shows
$R-K$ against redshift for three models, computed using Pegase2 (Fioc
and Rocca-Volmerange 1997) with the time-redshift relation for
$\Omega_m=0.3$, $\Omega_\Lambda=0.7$ and $h_{100}=0.55$.

(i) An evolving model representing an
 elliptical galaxy, which 
forms stars in a 1
Gyr dust-reddened burst at $z>3.4$. 
 The SFR and dust then fall to zero, and 
 passive evolution occurs. The galaxy would have 
$R-K>5.0$ at all $z>0.93$ and hence is a `pERO'.

(i) A pure,
non-evolving starburst, with very strong dust reddening of
$E(B-V)=1.0$ mag. Such `dsfEROs' are reddest
at $1<z<1.6$, but if as dusty as this model can be EROs at all
$0.5<z<3$.
  Starburst
galaxies with an old stellar component, or post-starbursts,
 could be EROs
with somewhat less reddening.

(iii) For comparison, a model representing an evolving spiral (Sbc) with 
near-solar metallicity and
a `normal' amount of
dust.  `Normal' star-forming spirals like this
 do not have  $R-K>5$ at any redshift.

\begin{figure}
\psfig{file=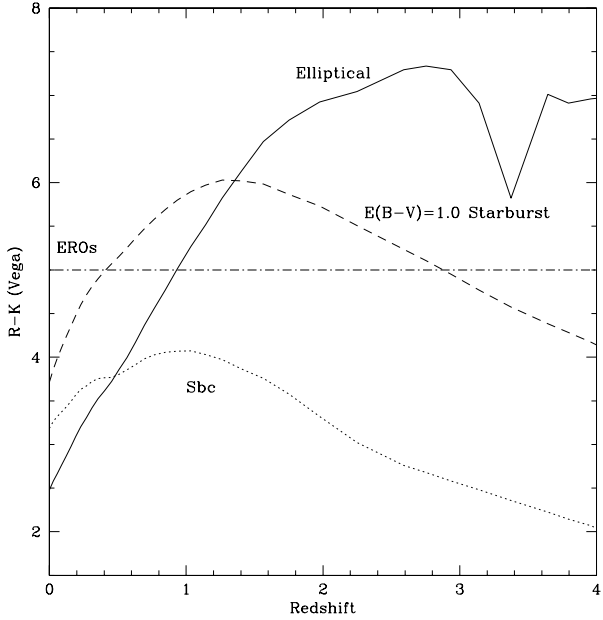,width=90mm}
\caption{Observed $R-K$ colour against redshift, for three models
representing an evolving
elliptical galaxy (pERO), 
 a heavily reddened starburst (dsfERO), and a normal spiral (which is never an ERO).}
\end{figure}
\subsection{ERO Selection}
The $K$-detected galaxies were
matched in RA and Dec with detections on the WHT $R$-band image,
within a maximum offset radius of 1.5 arcsec (two
small areas of Ingrid data were not covered in $R$, slightly reducing
the $K+R$ overlap region 
to 81.5 $\rm arcmin^2$). 
 The colour of
each galaxy was taken to be simply the difference of the $R$ and $K$
total magnitudes. To test this, the `total mag' colours were compared
with $R-K$ measured in
fixed circular (2.5 arcsec) apertures, for a subsample of EROs. These
colours agreed within $\leq 0.15$ mag in most cases, with no
significant systematic difference.

Any $K\leq 21$ detection with  
either $R-K>5$ or
no $R$ counterpart was flagged as a candidate ERO.
 The $K$ detections in the latter category are either 
 (i) spurious, (ii) real
galaxies where the $R$ counterpart is obscured (e.g. by a
diffraction spike) or merged with a brighter detection, (iii) real
galaxies too faint in $R$ to be detected on our image, which can 
reasonably be classified as  EROs. All were
carefully examined by eye, on the $K$ image and at the corresponding
position on the $R$ image. In this way, a number of spurious detections
were identified, and thereafter excluded.

 Our
final catalog of verified EROs consists of 158 objects (including 49
undetected in $R$), of which 
99 are on the UFTI mosaic.
Figure 2 shows $R-K$ against $K$ magnitude for the detected galaxies --
 EROs appear only at faint magnitudes, mostly $K>18$. 
One much brighter ($K=16.2$) object has $R-K=5.72$,
but appears stellar and is almost certainly a red 
 Galactic star, and is not
included here as an ERO.

\begin{figure}
\psfig{file=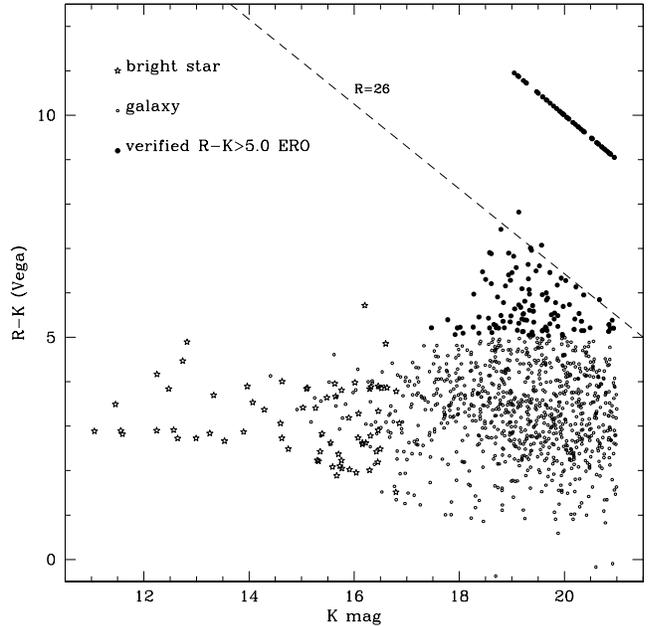,width=90mm}
\caption{$R-K$ colour against $K$ for 
UFTI and Ingrid  detections to $K=21$, with star-galaxy
separation to $K=16.5$. Galaxies undetected in $R$, and hence
included as EROs, are placed at $R=30$, but may lie anywhere above the 
$R=26$ locus.}
\end{figure}

 \subsection{Spatial Distribution}
Figure 3  shows the distribution of EROs.
ERO maps are useful for identifying galaxy clusters at $z\geq 1$,
where the early-type members will show up as an 
arcmin-scale overdensity 
(e.g. Stanford et al. 1997). In our data,
there may be a group of EROs centered on UFTI detection number 608, a bright
ERO ($K=17.78$) of particular
interest in that it is a radio and X-ray source (Section 7). Within
a 45 arcsec radius (shown on Figure 3) of its position 
there are 8 EROs with 
$K\leq 20$,
compared to 2.4 expected for a random distribution. These EROs
are discussed further in  Section 6.3.
\begin{figure}
\psfig{file=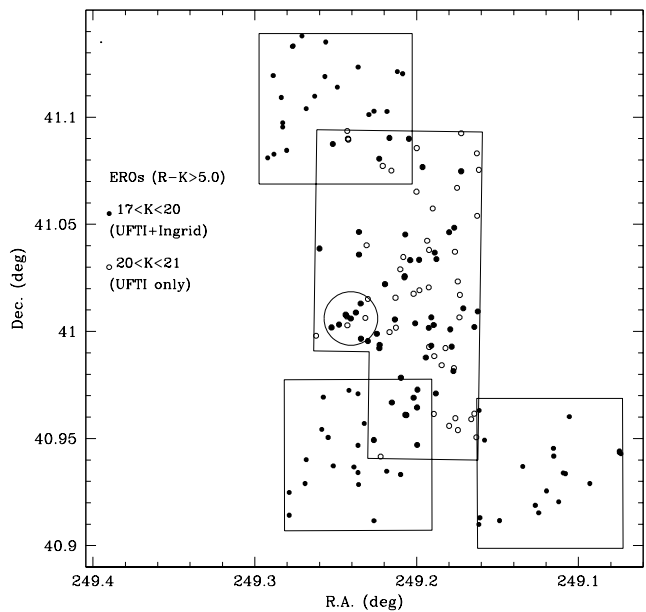,width=90mm}
\caption{Distribution on the sky of the $K\leq
20$ EROs ($R-K>5.0$) on the UFTI mosaic (central area) and the 3 Ingrid fields, with 
the $20<K<21$ EROs on  the UFTI field only. 
The plotted circle shows a 45 arcsec radius around UFTI detection 608.}
\end{figure}

\section{Number Counts of Galaxies and EROs}
\subsection{Observations}
Figure 4 shows $K$-band differential number counts for all 
 galaxies, from the
combined  UFTI+Ingrid data at $K\leq 20$ and UFTI data only at $20<K<21$.
Also shown are counts from the
deeper (to $K\sim 23$) Keck and ESO-VLT  surveys  of Moustakas et
al. (1997) and Saracco et al. (2001).
 Our counts are    reasonably consistent with these, although
lower at $K\geq 20$,
 suggesting some incompleteness -- the ratio of our  
galaxy count to that of Saracco et al. (2001) falls from 0.968 at
$19.5<K<20.0$ to 0.725 at $20.0<K<20.5$ and 0.611 at $20.5<K<21.0$.

\begin{figure}
\psfig{file=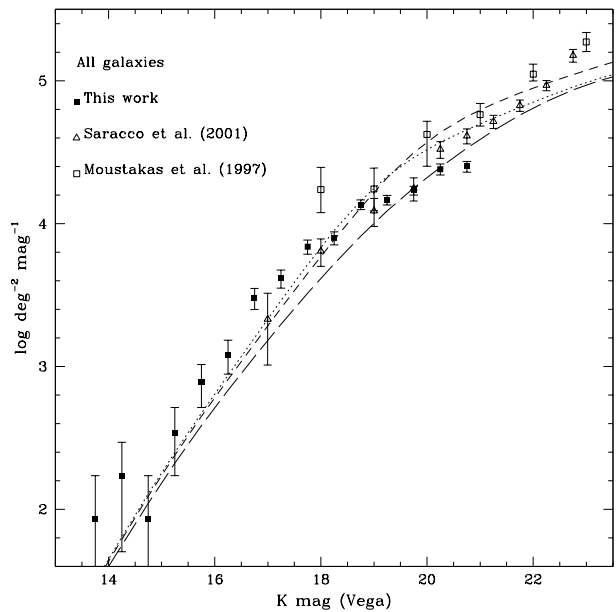,width=90mm}
\caption{$K$-band differential galaxy number counts, for galaxies of
all colours, from our data (Ingrid+UFTI to $K=20$, UFTI only at
$20<K<21$),  and 
the deeper surveys of Moustakas et al. (1997) and
Saracco et al. (2001), compared with three
 models; non-evolving (long-dashed), PLE (dotted), and
merging with $R_{\phi}=R_{m^*}=0.3$ (short-dashed).}
\end{figure}
Figure 5 and Table 1 show number counts for EROs only. The fraction of EROs in
the UFTI+Ingrid data is 112/812 =
13.8 per cent at $K\leq 20$ and 158/1076 = 14.7 per cent at $K\leq
21.0$. The `incompleteness corrected' count is derived by dividing
our observed ERO count by the ratio of our count for all 
galaxies 
to that of Saracco et al. (2001) at the same $K$ limit. Also plotted are
the $R-K>5$ ERO counts from Daddi et al. (2000), and the Saracco et
al. (2001)
ERO counts  -- note that the latter
 are selected with a  different criterion of $J-K>1.9$, 
which may include a greater number of `dsfEROs'.
Our ERO counts agree well with these two surveys.
\begin{figure}
\psfig{file=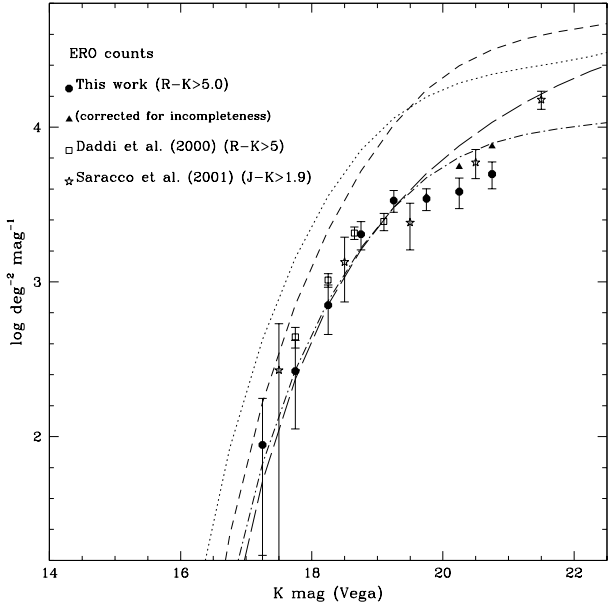,width=90mm}
\caption{$K$-band differential galaxy number counts for 
$R-K>5.0$ EROs from our UFTI+Ingrid data and Daddi et
al. (2000), and for $J-K>1.9$ EROs in
Saracco et al. (2001), with minimal $\surd N$ errors,
 compared with  $R-K>5.0$ galaxy counts
predicted by PLE (dotted), non-evolving (long-dashed),
merging with $R_{\phi}=R_{m}=0.3$ (short-dashed), and `M-DE' merging with
$R_{m}=0.3$ $R_{\phi}=-0.46$ (dot-dashed) models.}
\end{figure}
\begin{table}
\caption{Observed number counts (number $N_g$ and surface density, $\rho$) of
$R-K>5.0$ EROs on the UFTI+Ingrid field, with  $\surd N$
errors, and   (at $K>19.5$) $\rho$ with an 
estimated incompleteness correction.}
\begin{tabular}{lccc}
\hline
$K$  & $N_g$  & $\rho$ (Obs.) & $\rho$
(`Corrected') \\
\smallskip
interval &  & $\rm deg^{-2}mag^{-1}$ &  $\rm deg^{-2}mag^{-1}$\\
17.0--17.5  &   1    &  $88.4\pm 88.4$ & - \\
17.5--18.0 &    3   &  $256.0\pm 153.0$ & - \\
18.0--18.5 &   8    & $706.8\pm 249.9$ & - \\ 
18.5--19.0 &   23  &  $2032.2\pm 423.7$ &  - \\
19.0--19.5 &  38  &  $3357.5\pm 544.7$ & - \\
19.5--20.0 &  39  &  $3445.9\pm 551.8$ & 3560.  \\
20.0--20.5 &  20  &  $3827.2\pm  855.8$ & 5279.  \\
20.5--21.0 &  26  &  $4975.4\pm 975.8$ & 8143. \\
\hline
\end{tabular}
\end{table}

\subsection{Models: PLE, NE, Merging and `M-DE'}
These counts are compared with simple galaxy evolution
models, with and without the effects of merging. For the models
considered in this paper, we adopt cosmological parameters of 
$\Omega_M=0.3$, $\Omega_{\Lambda}=0.7$ (using the analytic luminosity
distance formula of Pen 1999) and the obseravtional (from 2MASS) 
$K$-band galaxy luminosity functions 
 of Kochanek et al. (2001). 
 luminosity evolution is modelled using Pegase2 (Fioc and
Rocca-Volmerange 1997) with a range of star-formation
histories, and (as in Roche et al. 2002), an initial mass
function with the Salpeter slope $x=2.35$ (where 
${d{\rm N}\over d {\rm M}}\propto \rm M^{-x}$) at $\rm
0.7<M<120~M_{\odot}$, 
flattening to $x=1.3$ at $\rm 0.1<M<0.7~M_{\odot}$.

 Galaxies in the early-type luminosity function
(represented as a Schechter function with 
$M^*_K=-25.04$, $\alpha=-0.92$ $\phi^*=0.0005625$ $\rm Mpc^{-3}$ for
$h_{50}=1$) form all their stars in an initial dust-reddened
($E(B-V)=0.65$ mag ) burst,
beginning 16 Gyr ago ($z\simeq 6$), with a range of   
durations represented as either 1 Gyr (for half) or 2 Gyr.
 After the burst both dust and SFR fall to zero
 and evolution is thereafter passive.
 Galaxies in the late-type luminosity function
($M^*_K=-24.49$, $\alpha=-0.87$ $\phi^*=0.0012625$ $\rm Mpc^{-3}$ for
$h_{50}=1$) are represented  with a range of 
continuous star-formation models with different timescales and
moderate dust (in the amounts calculated by Pegase). The only
$R-K>5$ galaxies in this model are the E/S0s at $z>0.93$.

In the pure luminosity evolution (PLE) model the galaxies evolve only
in $L^*$, with no change in $\phi^*$ or $\alpha$. As our version of
this model gives relatively strong $L^*$ evolution, we also show a 
 non-evolving (NE) model, with the same luminosity function  at
$z=0$. In  our merging model, the  evolution of luminosity per unit
mass in the same as the PLE model, but
superimposed on this is an  increase in comoving number density
($\phi^*$) 
and an associated decrease in the characteristic galaxy mass
($m^*$) with redshift, cancelling out some of the evolution in galaxy 
luminosity ($L^*$).

The merger rate is parameterized in terms of the 
 effect of merging on galaxy mass,  ${\Delta(m^*)\over m^*}=
R_m(z)$ per
Hubble time  ($t_H$). Observationally, 
Patton et al. (2001) estimated from CNOC2 data that  $R_m(z)\simeq
0.3(1+z)^{2.3}$ to $z= 0.55$.
 If we assume this evolution  to continue to  $z=1$, with $R_m(z)$ constant 
at $z\geq 1$, then $m^*(z)$ will evolve as:
$${dm^*\over dt}=m^*_z R_m(0)~t_H^{-1}(1+z)^{2.3}~~~{\rm  at}~z<1$$
$${dm^*\over dt}=m^*_z R_m(0)~t_H^{-1}2^{2.3}~~~{\rm at}~z\geq 1$$
 approximating the lookback time $t_{now}-t\simeq t_H-t_H(1+z)^{-1}$,
gives ${dt\over dz}=-t_H(1+z)^{-2}$. Substituting,
$${dm^*\over dz}=-m^*_z R_m(0)~t_H^{-1}(1+z)^{0.3}~~~{\rm at}~z<1$$
$${dm^*\over dz}=-m^*_z R_m(0)~t_H^{-1}2^{2.3}(1+z)^{-2}~~~{\rm at}~z\geq 1$$
Integrating,
$$\int^z_0 {dm^*\over dz}=R_m(0)\int^z_0 (1+z)^{0.3}dz$$
$${\rm ln}~m^*(z)={\rm ln}~m^*(0)-R_m(0)((1+z)^{1.3}-1)/1.3$$
$$m^*(z)=m^*(0~{\rm exp}~[-R_m(0)((1+z)^{1.3}-1)/1.3]~~~{\rm}~z<1$$
and,
$$\int^z_1 {dm^*\over dz}=2^{2.3} R_m(0)\int^z_1 (1+z)^{-2}dz $$
$${\rm ln}~m^*(z)={\rm ln}~m^*(1)-2^{2.3} R_m(0)({1\over 2}-
(1+z)^{-1})$$
$$m^*(z)=m^*(1)~{\rm exp}~[-2^{1.3} R_m(0)(1-2(1+z)^{-1})]~~~{\rm 
at}~z\geq 1$$
with $R_m(0)=0.3$ this gives $m^*(1,2,3)=
(0.714, 0.558, 0.493)m^*(0)$. If the corresponding opposite evolution in
comoving number density, $R_{\phi}$, occurs at the same rate
($R_{\phi}=R_m$),
  galaxies at $z\sim 3$ are half as
bright but twice as numerous as they would be in a PLE model.

 The counts of all galaxies (Figure 4) exceed the NE model, and are
 more consistent with PLE or merging. The observations do not firmly
distinguish between these two, as  
the model counts are
relatively insensitive to merging. Beyond the limits of our survey
 there is
 some indication that the count exceeds all three  models, which
 may indicate the  luminosity function has a 
steeper faint-end slope 
 than derived by Kochanek et al. (2001).

 On Figure 5 the
same models are used to predict the counts of $R-K>5.0$ EROs.
Although only the `pERO' type is included, 
our PLE model overpredicts the ERO counts at all magnitudes, by
factors $\sim 3$. The inclusion of $R_{\phi}=R_m=0.3$ merging improves the
fit at the bright end, but increases the overprediction at $K>20$.
However, the non-evolving model does fit the ERO counts.

Firth et al. (2002) similarly found number counts of EROs to be
well-fitted by a non-evolving model to $H=20$ ($K\simeq 21$), and also
claim that this model fits their photometrically-estimated redshift 
distribution $N(z)$. Our PLE model, with its relatively top-heavy IMF
and ejection of the dust at the end of the starburst, gives 
 strong $L^*$ evolution for
E/S0s, and in combination with the high formation redshift,
predicts a high ERO count. `Milder' PLE models may be somewhat closer to the
observed ERO counts, but to be consistent with them, we and Firth et
al. (2002) had to
reduce $L^*$ evolution to near-zero. Cimatti (2000b) did fit the ERO
counts with a PLE model, but only for an elliptical
formation redshift  $z\leq 2.2$, which is unrealistically low  in view of the
existence of Lyman break objects, SCUBA sources and radio galaxies at
$z>3$. Hence we conclude  that the number counts of EROs
 are in
general inconsistent with PLE models.
     
With the aim of finding a physically realistic evolving model for ERO
counts, we consider the  merging model again, 
and retain  $R_m=0.3$ while varying $R_{\phi}$ to best-fit our 
incompleteness-corrected ERO count. The minimum  
$\chi^2$ occurs  for $R_{\phi}\simeq -0.46$ (with $1\sigma$ error bar
$\pm0.10$). Figure 5 shows the counts for this model, hereafter
`M-DE' (merging with negative density evolution). They are very similar to
the NE model counts, to $K\simeq 20$, and in good agreement with
observations.

 Figure 6 illustrates the features of the M-DE
model. Both mass and comoving number density decrease with redshift. 
Luminosity per unit mass (shown for restframe $R$ and $K$) increases
with redshift due to passive evolution (at $z>3.4$ the galaxy is
starbursting and most of the red light is hidden by dust). Luminosity
($L^*$) increases with $z$ at a slower rate as it is partially
cancelled
by the decrease in mass.
 Physically, this model could be
represented as 60 percent of the present-day comoving
number density of  E/S0 galaxies
forming at
$z>1$, and 42
 percent at $z>2$, with the remainder originating at
lower redshifts, e.g. from mergers of disk galaxies in
which the remaining gas content is exhausted in a
 merger-triggered starburst. Some spiral-to-elliptical transformation  could
also occur through the  cumulative effect
of non-merging interactions, or gas-stripping in clusters.

 \begin{figure}
\psfig{file=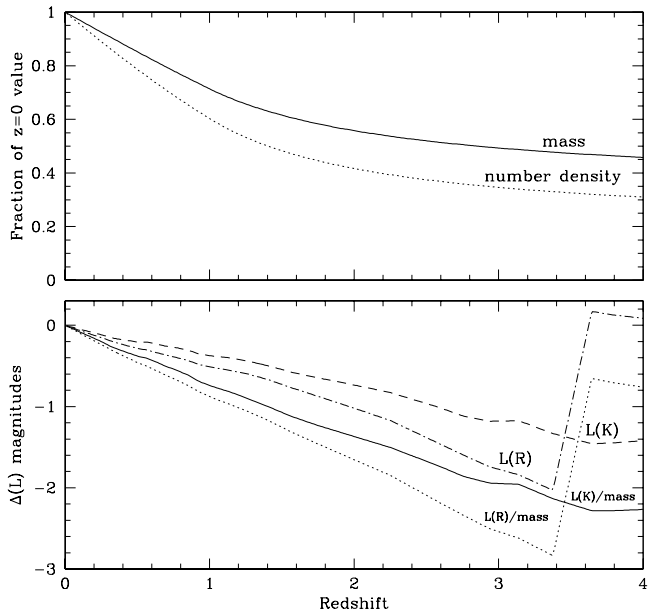,width=90mm}
\caption{Features of the `M-DE' model; (above) evolution of
characteristic galaxy mass $m^*$ and comoving number density $\phi^*$;
(below) evolution of the characteristic luminosity $L^*$, in the
rest-frame $K$ and $R$ bands, and of the luminosity per unit mass, for
an elliptical with a 1 Gyr initial starburst (star-formation at
$3.4<z<6$).}
\end{figure}

In addition to rejecting the PLE model, the
ERO counts do not support the other extreme case of
`pure' hierarchical merging. Cimatti (2002b) compares the
observed count of passive EROs at $K\leq 19.2$ with the predictions
of two such models from Firth et al. (2002) and Smith et al. (2002),
both derived from $\Lambda$CDM cosmologies. These models predict,
respectively, 0.19 and 0.04 EROs $\rm arcmin^{-2}$ at $K\leq 19.2$.
Our observed (UFTI+Ingrid) count of EROs to this limit is $0.60\pm
0.09$  $\rm arcmin^{-2}$, and hence a 
mixture of formation histories, as represented by the M-DE model, is
preferred.

\section{Clustering of EROs}
\subsection{Calculating the  Angular Correlation Function}
The clustering properties of the EROs will provide clues to their
nature and history. We investigate the clustering 
by  calculating the angular correlation function, $\omega(\theta)$,
 for both $R-K>5$ galaxies
and full $K$-limited samples, by the methods  described below. 
To $K=20$ we use the combined UFTI and Ingrid catalogs, but for fainter
limits the analysis is limited to galaxies on the  UFTI mosaic.
 
 For $N_{g}$ 
galaxies brighter than a chosen magnitude limit, there will be
${1\over2}N_{g}(N_{g}-1)$ possible galaxy-galaxy pairs. These are 
counted in bins of separation of width $\Delta ({\rm log}~\theta)=0.2$,
giving a function $N_{gg}(\theta_i)$.
A large number of random points ($N_r=50000$ here)
is scattered  over the same area as covered by the real galaxies, and
 the separations of the $N_{g}N_{r}$ galaxy-random pairs, taking the
real galaxies as the centres, are  counted in bins to give $N_{gr}(\theta_i)$. 
The separations of the ${1\over2}N_r(N_r-1)$ random-random
pairs are also counted to give $N_{rr}(\theta_i)$. 

If $DD=N_{gg}(\theta_i)$, and $DR$ and $RR$ are the galaxy-random
and random-random counts normalized to have the same summation over $\theta$,
$$DR={(N_g-1)\over 2N_r} N_{gr}(\theta_i)$$
$$RR={N_g(N_g-1)\over N_R(N_r-1)} N_{rr}(\theta_i)$$
then, using the  Landy and Szalay (1993) estimator,
$$\omega(\theta_i)={DD-2DR+RR\over RR}$$
Errors were estimated by dividing the data area into 20 sub-areas and then 
recalculating $\omega(\theta)$ for 20 subsamples, each time excluding both the
real galaxies and the random points from a different sub-area and
using the remaining 19. The scatter between the $\omega(\theta)$ of
these 20 subsamples is then
multiplied by $\surd {19 \times 19\over 20}=4.25$ to give 
the error bars for the full dataset  $\omega(\theta)$. 
 
This estimate will be negatively offset from the true $\omega(\theta)$ due to
the restricted area of observation (the `integral constraint'). 
If the real $\omega(\theta)$ is of the form
$A_{\omega}\theta^{-\delta}$, where $A_{\omega}$ is an amplitude, the
 estimate corresponds to $A_{\omega}(\theta^{-\delta} - C)$. The
negative offset $AC$ can be estimated by doubly
integrating an assumed true $\omega(\theta)$ over the field area $\Omega$,   
$$AC={1\over \Omega^2}\int\int \omega(\theta) d\Omega_1 d\Omega_2$$
Using the random-random
correlation, this  can be done numerically --
$$C={\sum N_{rr}(\theta) \theta^{-\delta}\over \Sigma N_{rr}(\theta)}$$
Assuming $\delta=0.8$, this gives  $C=9.226$ for the combined
UFTI+Ingrid 
area, and 13.46 for the UFTI mosaic only.

\begin{figure}
\psfig{file=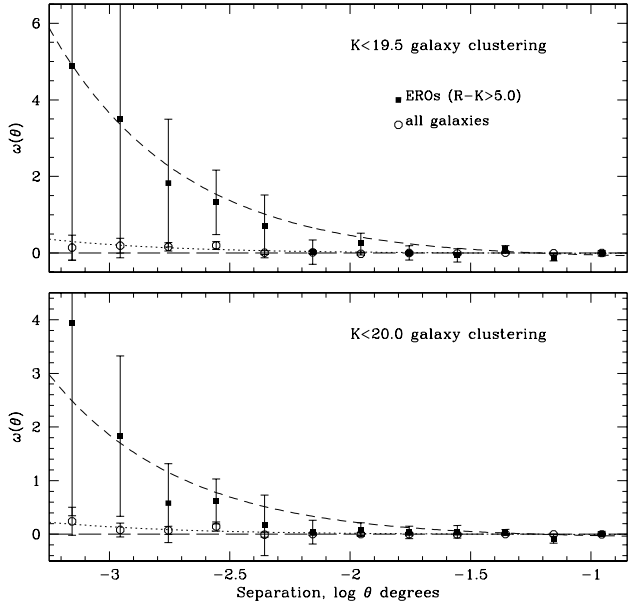,width=90mm}
\caption{Angular correlation functions, $\omega(\theta)$ as calculated
for $R-K>5.0$ EROs, and for all galaxies, on the combined 
UFTI+Ingrid  field to limits of (a) $K\leq 19.5$ and (b) $K\leq 20.0$, with best-fit functions of the form
$`A_{\omega}(\theta^{-0.8}-C)'$
(dotted).}
\end{figure}

 The amplitude 
$A_{\omega}$ is then obtained by least-squares fitting
$A(\theta^{-0.8}-C)$ to the observed  $\omega(\theta)$, over the
range $1.26< \theta < 502$ arcsec, weighting
each point using the error bars estimated as above.
To estimate an  error on $A_{\omega}$, the same 
function is then fitted 
to the $\omega(\theta)$ of each of the 20 subsamples, 
 and multiplying the scatter 
between the subsample $A_{\omega}$ by, again,  4.25.

\subsection{$\omega(\theta)$ results}
For the EROs we detect positive clustering at the $>2\sigma$ level at
magnitude limits $K=19.5$--20.0. To $K=20.25$ the  smaller sample
provided by the only the UFTI data gives only $1.5\sigma$ evidence of
clustering. At $K=20.5$--21.0, there is no detection even at
$1\sigma$, but we do not consider these limits here as
$\omega(\theta)$ is likely to be
affected by the incompleteness. 
For the full $K$-limited sample the detection of clustering never
reaches $2\sigma$, and the upper limits indicate the $A_\omega$ is
about an order of magnitude lower than for EROs at the same $K$ limits. 
 
Figure 7 shows the observed $\omega(\theta)$, with the fitted
functions, at two $K$-band limits, and Table 2 gives the best-fit
$A_{\omega}$ with $\pm 1 \sigma$ errors. Figures 8 and 9 show
our $A_{\omega}$ against $K$ limit, together with results from
previous clustering analyses and models (described below).
In general our results are 
consistent with all the published $A_{\omega}$,
 support the strong clustering of EROs
reported by Daddi et al. (2000) and Firth et al. (2001), and suggest
that this continues to at least $K\simeq 20$. 
\begin{table}
\caption{Galaxy $\omega(\theta)$ amplitudes $A_{\omega}$ (in units of $10^{-4}$ at one
degree) of full $K$-limited galaxy samples and of
EROs ($R-K>5.0$), as estimated from  UFTI+Ingrid data (UFTI only at
$K>20$). $N_g$ is the number of
galaxies in each sample.}

\begin{tabular}{lccccc}
\hline
$K$ mag & \multispan{2} All galaxies & \multispan{2} $R-K>5.0$ EROs \\
\smallskip
limit & $N_g$  & $A_\omega$ & $N_g$ & $A_\omega$ \\
19.50 & 615 & $9.30\pm 5.26$ &  73 & $151.2\pm 59.2$   \\ 
19.75 & 714 & $5.32\pm 4.48$ &  93 & $86.63\pm 35.26$  \\  
20.00 & 813  & $5.76\pm 3.65$ &  112  & $76.61\pm 33.25$ \\
20.25 & 418 & $8.24\pm 4.93$ &  63 & $36.02\pm 24.29$  \\
\hline
\end{tabular}
\end{table}
\begin{figure}
\psfig{file=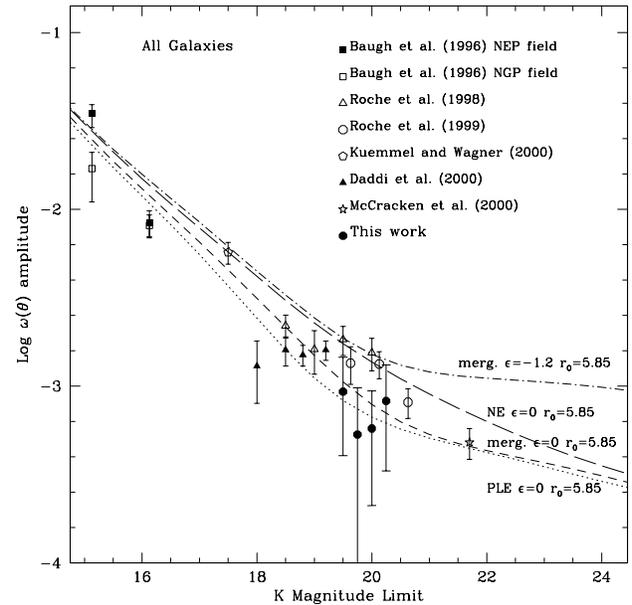,width=90mm}
\caption{The scaling of $\omega(\theta)$ amplitude with $K$ magnitude
limit for full $K$-limited samples of galaxies, as derived from our
data, shown  with previously
published results and four models (all with 
$\Omega_M=0.3$ $\Omega_{\Lambda}=0.7$ and $r_0=5.85 ~h_{100}^{-1}$ Mpc) -- 
PLE with $\epsilon=0$ (dotted), merging (with
$\epsilon=0$ (short-dash), NE with $\epsilon=0$ (long-dash)
 and M-DE with comoving ($\epsilon=-1.2$) clustering.
 (dot-dashed).}
\end{figure}
\begin{figure}
\psfig{file=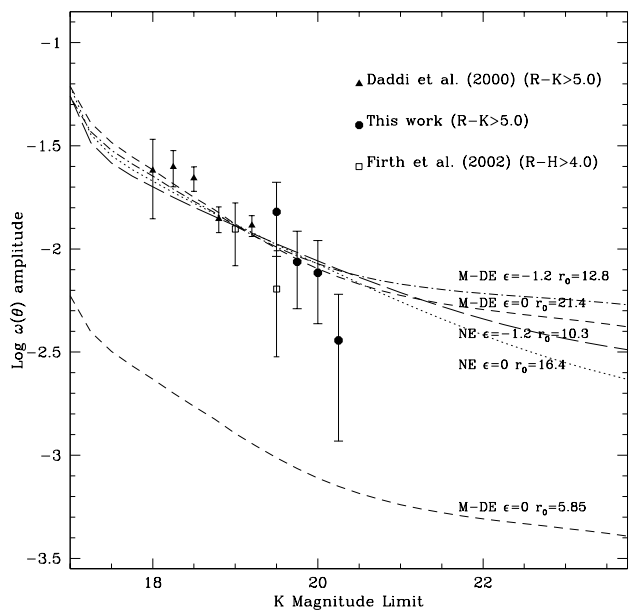,width=90mm}
\caption{As Figure 8 for EROs ($R-K>5.0$ galaxies) only. The $H$-band
limits of Firth et
al. (2002)  are converted to $K$  assuming
$H-K=1$ (approximately valid for $0.8<z<2.6$ ellipticals).
The plot shows four models fitted to the observed $\omega(\theta)$
scaling;
comoving M-DE with $r_0=12.8 ~h_{100}^{-1}$ Mpc (dot-dash), stable M-DE with
 $r_0=21.4 ~h_{100}^{-1}$ Mpc (long-dash), comoving NE with  $r_0=10.3
~h_{100}^{-1}$ Mpc (short-dash) and stable NE with $r_0=16.4
~h_{100}^{-1}$ Mpc. Also plotted, for comparison, is the M-DE stable
model with $r_0=5.85 ~h_{100}^{-1}$ Mpc, showing the order-of-magnitude
difference between ERO cluatering and that of `average' faint galaxies.} 
\end{figure} 

 \subsection{Modelling and Interpretation of $\omega(\theta)$} 
The observed $A_{\omega}$ of any sample of galaxies  depends on their
 redshift distribution
$N(z)$, and the cosmological geometry (through the angular size
distance), as well as on the intrinsic clustering of the galaxies, as
parameterized by their two-point correlation function in three
dimensions, 
$\xi(r,z)$. Hence to interpret $A_{\omega}$, we begin by modelling
$N(z)$. 

Figure 10 shows the $N(z)$ of EROs, at 
 $K\leq 19.5$ and $K\leq 21.0$, as given by our NE, PLE and M-DE
models.
PLE gives the highest numbers of EROs and most extended
 $N(z)$, while 
 the M-DE model $N(z)$ is much 
lower  in normalization and a little less extended in redshift, 
especially at the brighter limit. The NE model, as previously noted, 
 gives similar 
 numbers of EROs to M-DE, but the $N(z)$ is even less extended and peaks at
 a lower redshift.

The intrinsic (three-dimensional) 
clustering of galaxies and its evolution with redshift is represented here
by the power-law form
$$\xi(r,z)=(r/r_0)^{-\gamma}(1+z)^{-(3+\epsilon)}$$ 
where $r_0$ is the local correlation radius, $\gamma\simeq 1.8$ (observationally) and
$\epsilon$ represents the evolution with redshift, $\epsilon=0$ being
clustering stable in proper co-ordinates, and
$\epsilon=-1.2$ comoving clustering.
This produces a projected (two-dimensional) clustering,
$\omega(\theta)=A_{\omega}\theta^{-(\gamma-1)}$, where 
the amplitude $A_{\omega}$ 
is given by  Limber's formula (see e.g. Efstathiou et al. 1991),
$$A=C_{\gamma} r_0^{\gamma} \int_0^{\infty}{ (1+z)^{\gamma-(3+\epsilon)}\over
x^{\gamma-1}(z){dx(z)\over dz}}[(N(z)^2] dz/[\int_0^{\infty}N(z)dz]^2$$
where $x(z)$ is the proper distance 
and $C_{\gamma}=3.679$ for $\gamma=1.8$.

\begin{figure}
\psfig{file=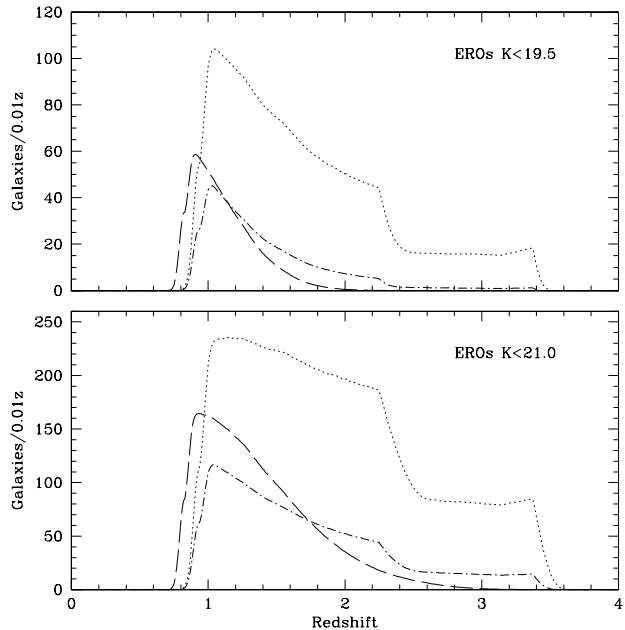,width=90mm}
\caption{The redshift distribution $N(z)$ of $R-K>5.0$
galaxies (pEROs) predicted by  
 PLE (dotted), NE (long-dashed) and M-DE (dot-dash) models, at limits $K\leq 19.5$ and
$K\leq 21.0$.}
\end{figure}
 Figure 8 shows four  models for the $\omega(\theta)$
scaling of all galaxies with $K$-band limit,  all with the
normalization $r_0=5.85
~h_{100}^{-1}$ Mpc, as estimated by Cabanac, de Lapparent and Hickson (2000)
 from an $I$-band survey. The
NE model predicts a higher $A_{\omega}$ than the  PLE and merging 
($R_{\phi}=R_{m^*}=0.3$) models, due to the less extended $N(z)$,
and a negative $\epsilon$ increases $A_{\omega}$.

 The observed $A_\omega$
are consistent with or slightly above the merging and PLE models with
 $\epsilon\simeq 0$. The NE and especially the comoving model
overpredict the $\omega(\theta)$ to some extent.
  
Figure 9 shows some models which fit the $A_{\omega}$ of EROs.
Giant ellipticals are found in local surveys 
to be more strongly clustered than `average' galaxies; e.g. 
 Guzzo et al. (1997)
estimate $r_0=8.35\pm 0.75  ~h_{100}^{-1}$ Mpc for E/S0s with
$M_B<-21$.
If the EROs, or most of them, are indeed 
the progenitors of massive ellipticals, they 
would be expected to show  similarly strong clustering. To estimate
the intrinsic clustering ($r_0$) of EROs, we very the $r_0$ to
 minimize  the $\chi^2$ for the 7 plotted data points (from
all three surveys) at
 $18.8\leq K\leq 20.0$, for our
M-DE and NE models, and  for both $\epsilon=0$ and
$\epsilon=-1.2$.

 For comoving clustering, we best-fit
$r_0=12.8
\pm 1.5  ~h_{100}^{-1}$ Mpc for the M-DE model and  $r_0=10.3\pm
1.2 ~h_{100}^{-1}$ Mpc for NE, and  for stable clustering,  
$r_0=21.4
\pm 2.6  ~h_{100}^{-1}$ Mpc for the M-DE model and  $r_0=16.4\pm
2.0 ~h_{100}^{-1}$ Mpc for NE. We note that 
Firth et al. (2002) estimate very similar $r_0$ for $H<20.0$ EROs and
a photometrically estimated $N(z)$. Our four fitted models are very
similar to $K=21$, and all are
consistent within $1\sigma$ of the observations -- much deeper data will be
needed to  constrain the evolution within the $-1.2<\epsilon<0$ range. 

 In summary,  for the probable
range of E/S0 luminosity evolution, we estimate  the intrinsic clustering
of  bright, $K\leq 20$, EROs, in the form of a  comoving correlation
radius, as 
$r_0\simeq 10$--13$
~h_{100}^{-1}$.
 This implies  that the intrinsic 
clustering of at  least the brighter 
EROs is indeed stronger than that of `average' galaxies, and even
exceeds the $r_0\sim 8 h^{-1}_{100}$
Mpc    of  present-day giant ellipticals, 
although it may not quite equal  the    $r_0\simeq 21 h^{-1}_{100}$
Mpc of local Abell clusters (Abadi, Lambas and
Muriel 1998). This is discussed further in Section 8.2.2. 
\onecolumn
\begin{table}
\caption{Co-ordinates (equinox 2000.0) of the 32 EROs in our bright
sample ($K\leq 19.5$ on UFTI image), $K$ and $R$ magnitudes,
best-fit $r_{hl}$ (in arcsec,
with error)  ellipticity $1-{b\over a}$ ($0=$ round),  best-fit profile type
($d=$ disk $b=$ bulge $p=$ point-source), significance 
 $\Delta(\chi^2)$, in units $\sigma(\chi^2)$  of favouring disk or
bulge,   point-source fraction $f_p$, and other
properties of 
morphology ($int$ = in interacting pair, $dn=$ double
nucleus, $asn=$ asymmetric -- offcentre nucleus, $aso=$ asymmetric outer
regions, $ir=$ irregular/flocculent)
) or emission ($X=$ Chandra x-ray detection, $Ra=$ VLA radio detection).}
\begin{tabular}{lcccccccccc}
\hline
\smallskip
Number & R.A & Dec.   & $K$ & $R$ & $r_{hl}$ & ell & Type  & sig & $f_p$ & other  \\
r91 & 16:36:54.32 &  40:56:57.43 &  19.084 &  25.655 &  $0.35\pm 0.10$
  & 0.188 & $b$ & 2.1 & 0 &  \\
r196 &   16:36:47.92 & 40:57:52.13 &  19.034 &  25.859 &  $0.18\pm  0.04$ 
   & 0.128 & $b$ & 1.03 & 0 & \\
r256 & 16:36:48.43 & 40:58:08.46 & 18.881 & 25.038 &  $0.38\pm 0.15$ &
 0.061 & $b+p$ & 2.1 & 0.375 & $Ra$ \\
r270 & 16:36:45.13 & 40:58:15.69  & 18.843 & 24.625 & $0.28\pm 0.04$ & 
 0.229 & $b$ & 4.0 & 0 &  \\
r280 & 16:36:47.87 & 40:58:22.07 & 19.325 &  24.376 & $0.20\pm 0.03$ & 
0.171 & $b$ & 1.05 & 0 & \\
r332 & 16:36:50.31 & 40:58:42.15 & 19.226 &  24.759 & $0.24\pm 0.03$ &
0.118 & $d$ & 0.47 & 0 & $asn$ \\
r407 & 16:36:46.62 & 40:59:16.15 & 19.472 & $>26$ & $0.78\pm 0.28$ &
0.106 & $d+p$ & 3.3 & 0.6 & $int$ \\
r492 & 16:36:53.51 & 40:59:31.85 & 19.309 & 25.951 & $0.19\pm 0.07$ &
0.019 & $b+p$ & 0.38 & 0.5 & \\
r506 &  16:36:42.77 & 40:59:34.36 & 19.410 &  24.997 & $0.31\pm 0.09$ &
0.090 & $b$ & 0.05 & 0 & $irr$ \\
r518 & 16:36:53.47 & 40:59:37.42 & 19.132 & $>26$ & $0.59\pm 0.18$ & 
0.230 & $b$ & 3.1 & 0 & $aso$ $Ra$ \\
r552 & 16:36:56.25 &  40:59:47.74 & 19.194 & 25.288 & $0.32\pm 0.04$ &
 0.039 & $b$ & 0.05 & 0 & $Ra$ \\
r561 & 16:36:53.89 & 40:59:55.92 & 18.019 & 23.246 &  0 &
 0.043 & $p$ & - & 1.0 & \\
r581 &  16:37:00.60 & 41:00:06.62 & 19.356 & 26.367 & $0.34\pm 0.03$ &
0.248 & $d$ & 0.10 & 0 & \\
r585 &  16:36:46.19 &  41:00:05.93 & 19.131 & 26.950 & $0.40\pm 0.10$ &
0.248 & $b$ & 1.67 & 0 & $asn$ \\
r594 & 16:36:39.43 & 41:00:07.48 & 19.377 & 26.338 & $0.32\pm  0.10$ &
0.210 &  $b+p$ & 0.81 & 0.25 & $aso$ \\
r599 & 16:36:45.46 &  41:00:10.81 & 19.463 &  25.957 & $0.42\pm 0.14$ &
0.062 & $b$ & 0.08 & 0 & $asn/aso$ \\
r608 &  16:36:57.76 & 41:00:21.71 & 17.780 & 23.176 & $0.50\pm  0.06$ &
0.152 & $b+p$ & 26.1 & 0.09 & $X$ $Ra$ \\
r622 & 16:36:45.80 & 41:00:23.72 & 18.501 &  24.802 & $0.66\pm  0.05$ &
0.325 & $d$ & 14.8 & 0 & $dn$ \\
r626 & 16:36:58.36 &  41:00:25.34 & 19.338 & 25.311 & $0.22\pm 0.07$ &
0.056 & $b$ & 0.67 & 0 & \\
r629 & 16:36:58.51 & 41:00:28.11 & 18.728 & 24.638 & $0.31\pm 0.05$ &
 0.104 & $b$ & 2.8 & 0 & \\
r642 & 16:36:56.99 & 41:00:31.76 &  18.610 & 24.819 & $0.32\pm  0.04$ &
0.105 & $b$ & 7.9 & 0 & $Ra$ \\
r650 &  16:36:38.92 & 41:00:33.62 & 18.984 & 24.466 & $0.59\pm 0.07$ &
0.438 & $d$  & 16.4 & 0 & $irr$ \\ 
r660 & 16:36:41.05 & 41:00:38.86 & 19.248 &  25.062 & $1.42\pm 0.30$ &
0.582 & $d$ & 12.9 & 0 & $irr$ $Ra$ \\
r675 & 16:36:56.29 & 41:00:46.77 & 18.695 &  23.960 & $0.61\pm 0.05$ &
0.429 & $d$ & $>20$ & 0 & \\
r852 &  16:36:49.75 & 41:01:32.89 & 19.222 &  25.575 & $0.28\pm 0.14$ &
0.173 & $b+p$ & 1.79 & 0.10 & \\
r952 &  16:36:45.31 & 41:02:12.34 & 19.111 & $>26$ & $0.57\pm 0.08$ &
0.351 & $d+p$ & $>2$  & 0.14 & $Ra$ \\    
r955 & 16:37:02.38 & 41:02:19.25 & 17.471 & 22.687 & $0.38\pm 0.03$ &
0.142  & $b$ & 17.8 & 0 & \\
r1091 & 16:36:49.68 & 41:02:42.80 & 19.404 & 24.454 & $0.45\pm 0.10$ &
0.057 & $d$ & $>10$ & 0 & \\
r1107 & 16:36:43.15 & 41:02:46.79 & 18.576 & 23.787 & $0.45\pm 0.05$ &
0.133 & $d+p$ & 0.9 & 0.12 & \\
r1114 & 16:36:56.54 & 41:02:47.07 & 19.105 & 24.265 & $0.39\pm 0.19$ & 
0.150 & $b+p$ & 0.4 & 0.50 & \\
r1127 & 16:36:42.39 & 41:02:54.31 & 18.376 &  23.831 & $0.50\pm 0.07$ &
0.270 & $d$ & 1.3 & 0 & $aso$ \\
r1437 & 16:36:53.52 & 41:04:50.34 & 19.283 & 24.667 &  $0.36\pm 0.06$ &
0.156 & $d$ & 0.02 & 0 & \\
\hline
\end{tabular}
\end{table}
\twocolumn
\section{Morphology and Radii of EROs}
\subsection{Profile Fitting and Classification}
We investigate the morphology and angular sizes of the EROs, to as
 faint a limit as possible for our ground-based data. 
 After
some experimentation, it was found that useful  morphological information 
could be extracted from the UFTI (but not Ingrid) images of galaxies to 
$K\sim 19.5$. Hence, this Section concentrates on the  subsample of 32 
 EROs on the UFTI mosaic with $K\leq 19.5$. 

Table 3 gives positions and  magnitudes of
these objects, labelled with their detection numbers in the
  SExtractor catalog (prefixed `r' for red). 
Using {\sevensize IRAF} `isophote.ellipse', a set of
elliptical isophotes were fitted to the UFTI images of each ERO, using 
centroids, ellipticities and position angles from the SExtractor
catalog as
the starting parameters. The surface brightness (SB) on each fitted
isophote, as  a function of semi-major axis, provides a radial intensity
profile.

For each ERO, we estimate a morphological classification (i.e. 
bulge, disk or point-source) and
 half-light radius, $r_{hl}$, but this is
 complicated by the need to correct for 
atmospheric effects. The seeing  point-spread function averaged 0.67
arcsec FWHM, but with significant variation between the 16 frames
of the UFTI mosaic. Hence to correct for this,    
 a bright (but not saturated), relatively  isolated, star was
identified on
the same frame as the each ERO.
 Using {\sevensize IRAF} `mkobjects',  a grid 
of model profiles was generated, including 

(i) a point-source,

(ii) a set of bulge profiles, 
$$I(r)=I_0~{\rm exp}[-7.67({r\over r_{hl}})^{-{1\over
4}}]$$

(iii) disk (exponential) profiles,
 $$I(r)=I_0~{\rm exp}-({r\over r_{exp}})$$ 
 covering a wide range of $r_{hl}$ in
steps of 0.01 arcsec (for a disk profile $r_{hl}=1.679 r_{exp}$). 
 This grid was convolved with the seeing point-spread function as
represented by the star (no sky noise is added at this stage). 
Using `isophote.ellipse', isophotal
profiles were extracted from each convolved model. Each is 
normalized to the same total intensity as the 
obseerved ERO profile, and $\chi^2$-tested against it. The model giving the
smallest $\chi^2$ is (after checking the fit by eye) then adopted as
the estimate for morphological type and seeing-corrected $r_{hl}$.

To estimate an error for $r_{hl}$, a second grid was generated,
containing multiple copies of the normalized, best-fit model
profile. This was convolved with the seeing, and noise equivalent
to the sky noise added. Isophotal profiles were
extracted from the resulting set of 
noisy models, and $\chi^2$ tested against the
observed ERO profile, and the 
scatter in the resulting $\chi^2$, $\sigma(\chi^2)$ calculated.
Returning to the first grid of
noiseless models, the change  $\Delta(r_{hl})$ away from the
 best-fit model that increases the $\chi^2$  by
$\sigma(\chi^2)$ is than taken as the $1 \sigma$ error on
$r_{hl}$.

Where possible both a disk
and a bulge model are fitted, so that
 the difference between their $\chi^2$, in
terms of  $\sigma(\chi^2)$, represents the significance by which one
profile is favoured.
Some EROs were not well fit by either model, but did give
a lower $\chi^2$ when a central
point-source component was included in the models.  For these, point
sources with flux fractions 
$f_p$ were combined with 
 bulge and disk models, and we determined the ($r_{hl}$, $f_p$) combination
which minimized $\chi^2$. In some other EROs, the 
 best-fit $\chi^2$ remained large because they have 
 `peculiar' morphologies, such as a double
nucleus. 

For each of the 32 bright EROs,
 the best-fit $r_{hl}$,  ellipticity (from
SExtractor),
best-fit  profile type
and the significance by which it is
favoured, are given in Table 3, and any obvious irregularity is noted. 
Figure 11 shows greyscale images  of 6 of these EROs, representing a
range of 
morphological types, and Figure 12 their
  radial intensity profiles with
best-fit models.
\begin{figure}
\vglue-10cm
\psfig{file=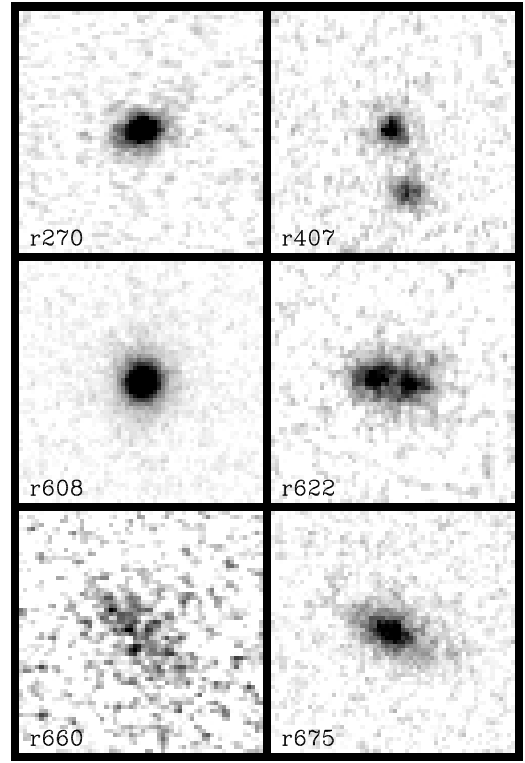,width=170mm}
\caption{Greyscale $K$-band (UFTI) images of 6 of the brighter EROs,
 showing $5.45\times 5.45$ arcsec areas. The galaxy
 r270 appears to be a normal elliptical, r407 a face-on
disk interacting with a fainter galaxy, r608 a
high-SB  elliptical at the centre of a possible cluster, r622 
 another merging system with a double nucleus, r660 
an extended, low-SB disk galaxy, and r675 an apparently regular  
spiral.} 
\end{figure}
\subsection{Morphology: Summary of Results}
(i) One of the 32 bright EROs, r561,  appears to be a pure point-source,
 and is probably a red Galactic star. 

(ii) Of the remaining 31 EROs, 19 were best-fitted with bulge and 
12 with exponential profiles. For some EROs the difference in
 $\chi^2$  between the two models is very small -- either they have  
 intermediate
profiles, and/or the signal/noise is insufficient for
classification. Of those where one profile is favoured by $>1\sigma$,
we classify  12  as bulges and 8 as disks.

(iii) A relatively small fraction of the EROs appear to be ongoing
mergers -- we find  2/31.
 A further 6 (4 bulge, 2 disk)
have some visible asymmetry,  most likely from recent
interactions.

(iv) The profiles of 9/31  EROs, including the `cluster' ERO, r608,
 were better fitted with the addition
 of central point-source
components to the underlying disk or bulge galaxies, 
although higher resolution data will be needed to confirm
 this.

The EROs at $K\leq 19.5$ appear to consist of 
about a 3:2 mixture of elliptical and spiral types, with about 1/4 
showing evidence of
ongoing or (more often) recent interactions. This is consistent 
 with the findings of
 Moriondo et al. (2001) and Stiavelli and Treu
(2001). 
\subsection{Angular Sizes of EROs}
Figure 13 shows the best-fit  half-light radii, $r_{hl}$,
 of the 31 (non-stellar) EROs 
against $K$ magnitude,
compared to models based on the radii of local galaxies. 
The size-luminosity relation of local E/S0s can be represented by the
double power-law relation of Binggeli, Sandage and Tarenghi (1984); for
 $M_B\leq -20$ (with $h_{50}=1$)
$${\rm log}~(r_{hl}/{\rm kpc})=-0.3(M_B +18.57)$$
and for $M_B> -20$ 
$${\rm log}~(r_{hl}/{\rm kpc})=-0.1(M_B +15.70)$$
An E/S0 galaxy with
$M_B=-21$ ($M_K=-25.04$) at $z=0$ would,  with the above size relation, 
have  $r_{hl}= 5.36$ $h_{50}^{-1}$ kpc. 
At $z=1$, with the luminosity evolution of our passive model, it
would have $K=18.22$, and an angular $r_{hl}$ of 0.48 arcsec.
 At $z=2$ it
would appear fainter, $K=19.44$ but little different in  size,
$r_{hl}=0.46$ arcsec. Figure 13 shows the whole $r_{hl}$--$M_B$
relation evolved to $z=1$ and 2.

\begin{figure} 
\psfig{file=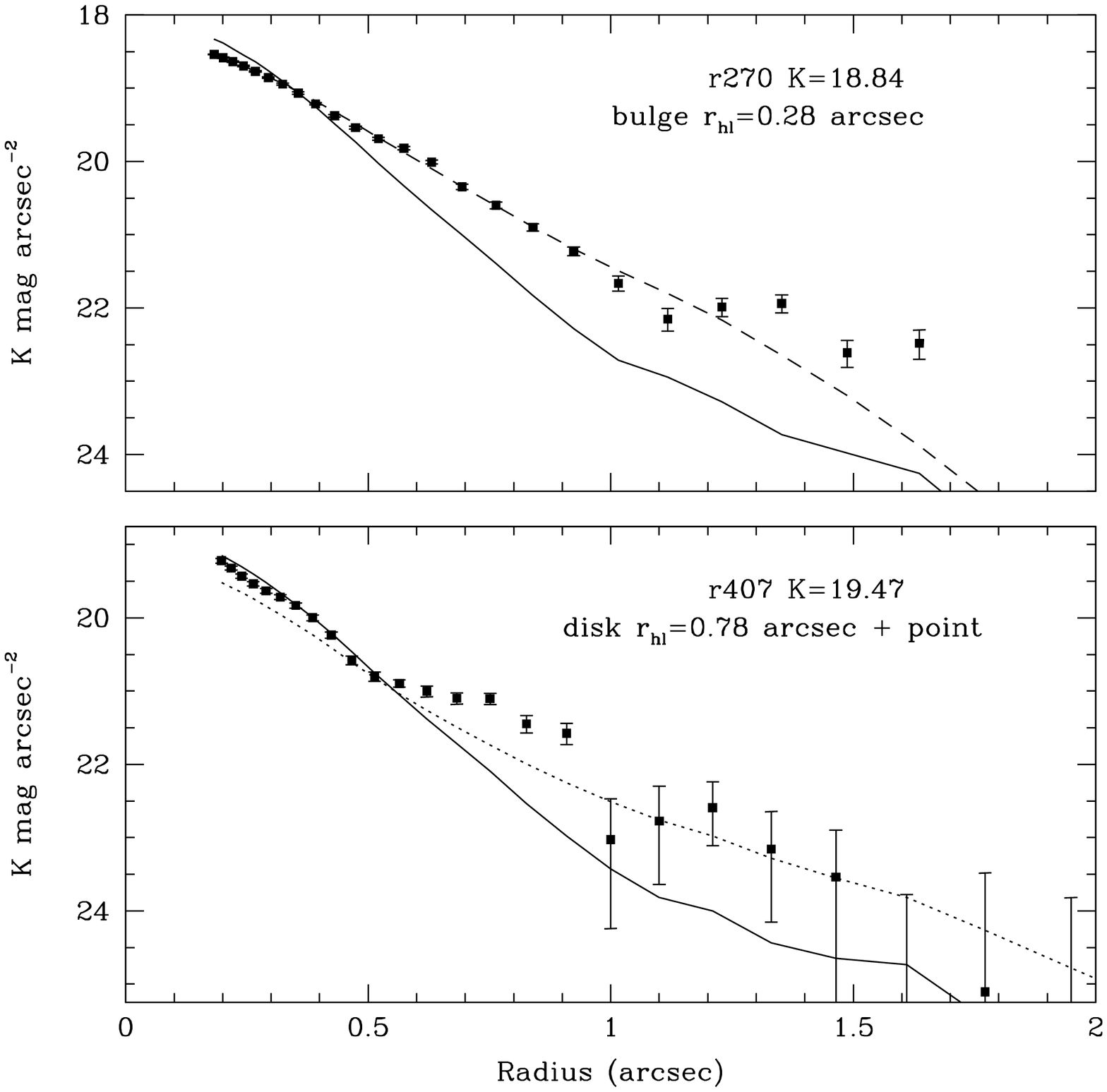,width=90mm}
\end{figure}
\begin{figure}
\psfig{file=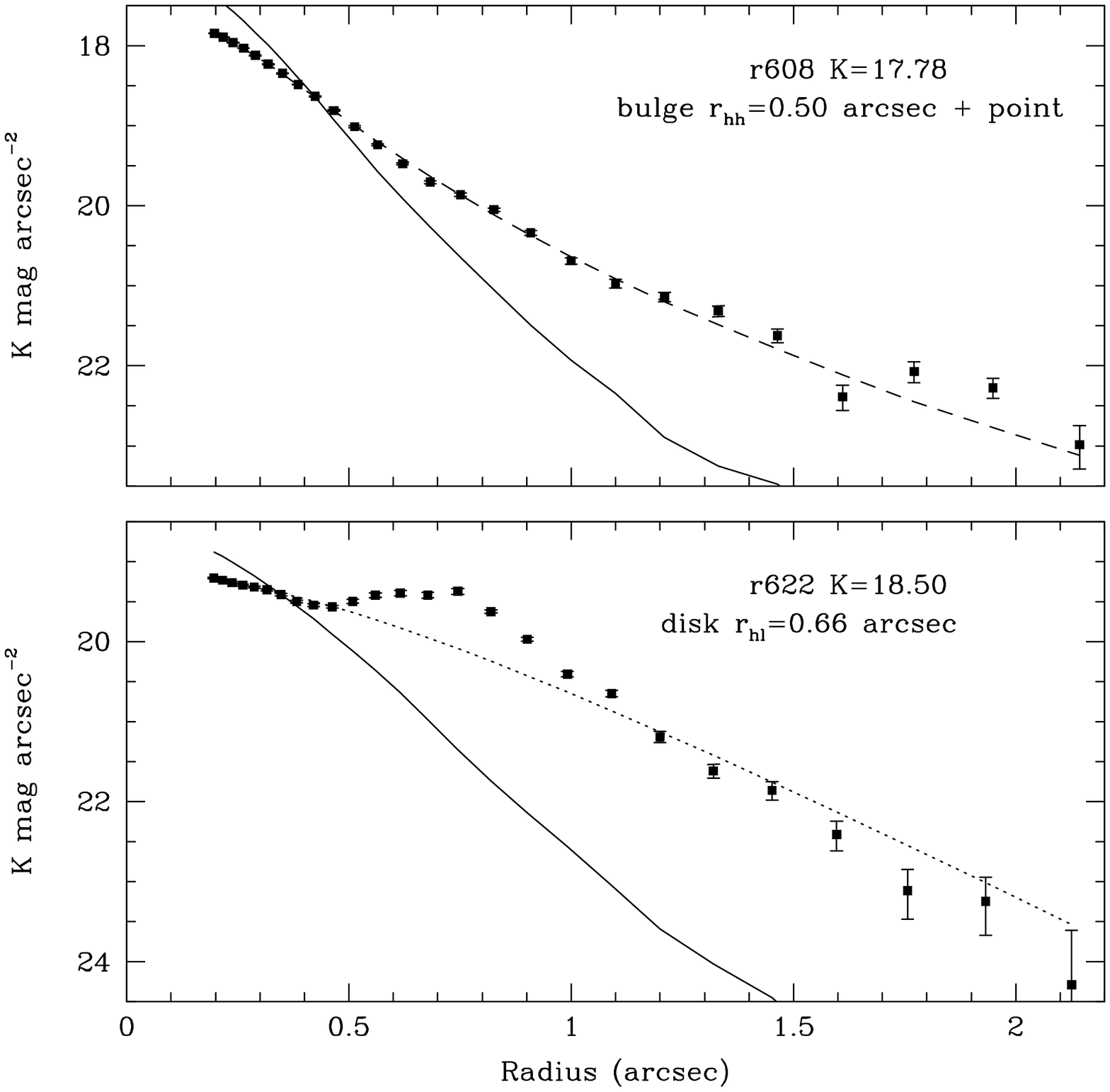,width=90mm}
\end{figure}
\begin{figure}
\psfig{file=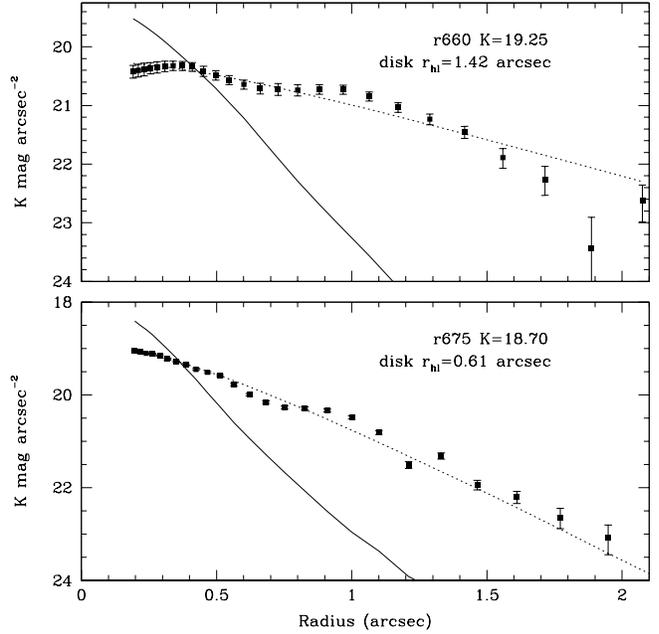,width=90mm}
\caption{Radial (major axis) intensity profiles of the six EROs shown
on Figure 11, compared with the best-fitting disk (dotted) or bulge
(dashed) model profiles, and the point-spread function (solid line)
 as determined from 
bright stars on the same CCD frames.}
\end{figure}

\begin{figure} 
\psfig{file=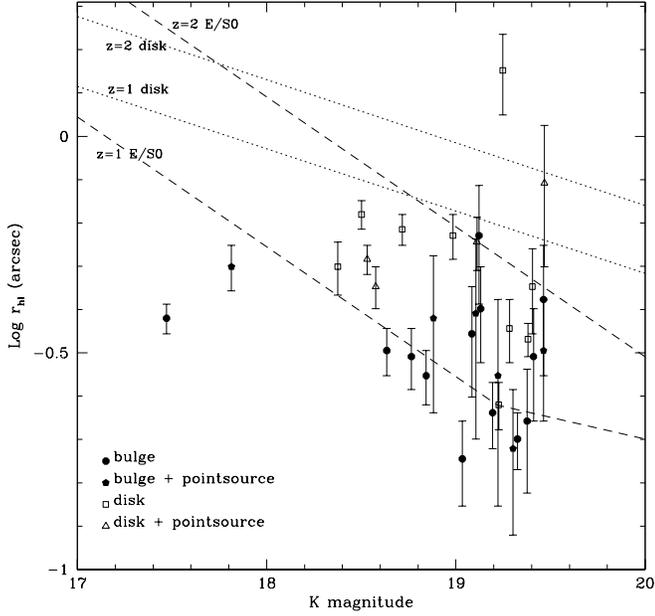,width=90mm}
\caption{Half-light  radii $r_{hl}$,  as estimated  by  fitting radial
profiles,  of the  32 UFTI-field EROs with $K\leq  19.5$ 
 (symbols  indicating profile type),  against $K$ magnitude -- compared  with
predicted  sizes for  E/S0 (dashed) and  disk (dotted)  galaxies at
$z=1$ and  2, based  on the size-luminosity  relations of  local E/S0s
(Binggeli , Sandage  and Tarenghi 1984) and spirals  (Cross and Driver
2002) with passive luminosity evolution.}
\end{figure}
The brightest of these EROs,  r955, has too high a surface brightness for
an elliptical at $z\sim 1$. As it is also only just red enough in 
to have been
included, the most likely explanation is that it is at a much
lower
 redshift. However, 
the other 18  bulge-type EROs  all
 have sizes consistent   with  passively  evolving
E/S0 galaxies at $0.9<z<2$.
 Most (12/19) are  concentrated near  the $z\sim 1$ locus
with only  two (r525, r612) on the    $z\sim 2$ relation, as would be
expected from the shape of $N(z)$ (Figure 10). 

Disk EROs are more likely to be dsfEROs, and hence their
 surface brightness may be increased or decreased
by starbursting and/or dust. 
Cross and Driver (2002) determine a bivarate brightness function for
 45000 disk galaxies in the 2dFGRS, which gives a  mean effective surface
 brightness $\mu_0=22.45$ $B$ mag $\rm arcsec^{-2}$ at $L^*$ ($M_B=-21.23$
for $h_{50}=1$), with a
positive correlation
 between surface brightness and luminosity. Their best-fit relation
 corresponds to
$${\rm log}~(r_{hl}/{\rm kpc})=-0.144M_B -2.034$$ 
with scatter $\sigma({\rm log}~r_{hl})=0.103$.
giving, for example $r_{hl}=9.77 h_{50}^{-1}$ kpc for $M_B=-21$,  hence
  $0.87(0.83)$ arcsec at $z=1(2)$.
A galaxy (of any morphology)
in which star-formation is truncated long before the epoch of
observation will be undergoing approximately the 
 passive (E/S0) 
model luminosity evolution, so with $M_B=-21$ locally would have
$K=18.22(19.44)$ at $z=1(2)$.

 Figure 13 shows the $r_{hl}-M_B$ loci of
this `passive disk' model at $z=1$ and 2.
Most (10/12) of the  disk  EROs  lie  below the $z=1$ locus,
so if they are at $z\geq 1$ their  intrinsic
 SB is greater than this model (some 
very dusty galaxies 
 may be included as EROs with
redshifts as low as $z\sim 0.5$ -- see Figure 1).
  The two exceptions are the interacting r407, which lies on the 
$z=2$ locus, and the very large r660, which is of a lower SB than this
model,
 suggesting it is strongly  dust-reddened (see Section 7).

Using the PLE and M-DE models, we
 predict $r_{hl}$ distributions for bulge
 and disk EROs. Figure 14 compares the $N(r_{hl})$  of the  $K\leq 19.5$
 bulge-type EROs in  our sample and Moriondo et al. (2000) with the models.
 PLE predicts  $\langle z \rangle
=1.72$ and $\langle r_{hl} \rangle =0.55$ arcsec for the bulge EROs,
 whweras with 
 M-DE they are a factor
 4.07 less numerous and have  $\langle z \rangle
=1.42$ and $\langle r_{hl} \rangle =0.43$ arcsec. With the NE
 model, the  EROs are just 4 per cent less fewwr than with M-DE, and
 havewith $\langle z \rangle
=1.11$ and $\langle r_{hl} \rangle =0.50$ arcsec For the 19 bulge EROs in
 our sample, 
$\langle r_{hl} \rangle =0.33\pm 0.03$ arcsec, but  the 16 of Moriondo et
 al. (2000) have  $\langle r_{hl} \rangle =0.49\pm 0.07$
 arcsec
(with most of the difference due to
two very large cluster ellipticals). The combined sample would  be
 reasonably consistent with  the M-DE model, but apparently
  inconsistent with NE
 which overpredicts the sizes.  
\begin{figure}
\psfig{file=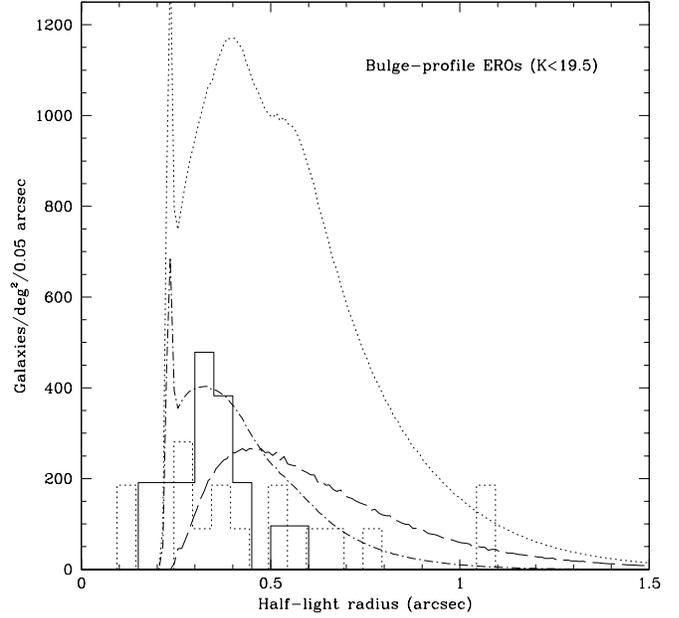,width=90mm}
\caption{Angular size distribution of $K\leq
19.5$ bulge-type EROs in our sample (solid histogram), and 
 Moriondo et
al. (2000) (dotted histogram), compared to
PLE (dotted), M-DE (dot-dash) and NE (long-dashed) models.}
\end{figure}

Figure 15 shows  model $N(r_{hl})$ for disk-type EROs, which adopt
 the spiral galaxy
$K$-band luminosity
function. Again the evolving models assume passive luminosity
 evolution.
 As our models do not predict the fraction of disk EROs we
have multiplied by arbitrary normalizations of
0.15 for PLE and 0.55 for M-DE and NE. These models  
are compared to histograms of the 12 disk-type EROs
in our sample, and the three  $K\leq
19.5$ EROs from Moriondo et al. (2000) that were classed as disks. 
\begin{figure}
\psfig{file=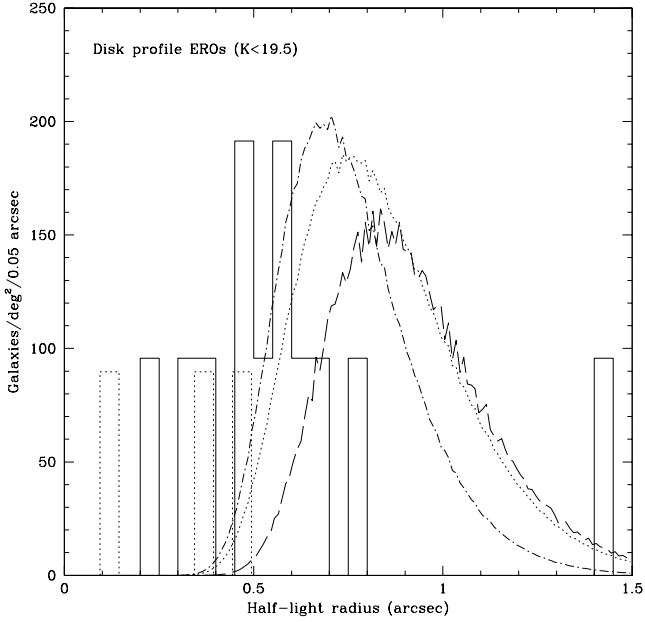,width=90mm}
\caption{Angular size distribution of $K\leq
19.5$ disk-type EROs in our sample (solid histogram), and Moriondo et
al. (2000) (dotted histogram); compared to
PLE (dotted) and   M-DE (dashed) models.}
\end{figure} 
The models predict   
$\langle r_{hl}\rangle=0.94$ arcsec for PLE, 0.76 arcsec for
M-DE, and 0.90 arcsec for NE.
In our sample the disk EROs have $\langle
r_{hl}\rangle=0.58\pm 0.09$.  Again the merging model is closest, although 
these disk EROs tend to be even more compact.   

If this size discrepancy is significant, it
could be the result of  size evolution --
there is some evidence that $z>1$ disk galaxies in general 
tend to be smaller in $r_{hl}$ than local counterparts of similar
mass (Roche et al. 1997, 1998; Giallongo et al. 2000).
Roche et al. (1998) hypothesised that this resulted from 
 `inside-outwards' disk formation, with
the star-formation becoming  more centrally concentrated at higher
redshifts. Disk $r_{hl}$ evolution appears to be quite
moderate ($\leq 0.1$ dex) at $z\sim 1$, but could be sufficient for
consistency with  the M-DE model. It is also
likely that 
 many EROs classed as disks are actually disturbed, post-interaction
 galaxies which may be the process of transforming into bulge
galaxies, and hence have   
 become more centrally concentrated than normal spirals.

\subsection{The EROs in a possible Cluster}
In Section 3.3 we noted a possible grouping of EROs centered on the
 X-ray and radio luminous r608. There are 11 $K<21$ EROs within a 45
 arcsec radius, of which 7
have $K\leq 19.5$  -- r552, r581, r608, r626,
 r629, r642, and r675, an overdensity  compared to the 1.6
$K\leq 19.5$ EROs expected by chance in this area.
If the EROs in this area belong to a single cluster, they
 should trace an iso-redshift locus on the $r_{hl}-K$ plot.
\begin{figure} 
\psfig{file=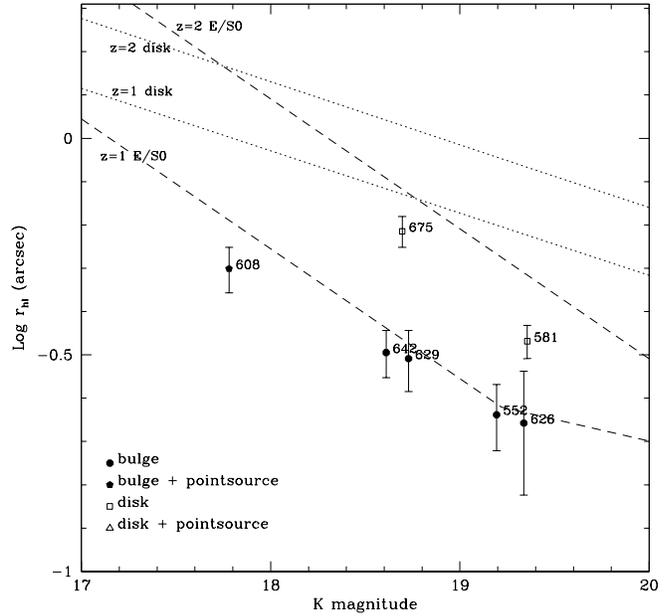,width=90mm}
\caption{Half-light  radii $r_{hl}$,  as estimated  by  fitting radial
profiles,  of the  7  EROs with $K\leq  19.5$ within a 45 arcsec
radius of r608, against $K$ magnitude. Galaxies are labelled with
detection numbers, with symbols  indicating profile type. The plotted
models are as in Figure 13.}
\end{figure}
On Figure 17, 
the five bulge-type EROs lie neatly on the $z=1$ E/S0 locus,
 and the two disks 
are 0.1--0.2 dex below the $z=1$ disk model. If
 these disks have a similar  size offset from this model as the
distribution on Figure 16, these radii are entirely consistent with
all 7 EROs belonging to a single, $z\sim 1$ cluster. Of course, to
confirm this will require 
spectroscopic redshifts (which we plan to obtain).

 \section{Radio, X-ray and sub-mm observations}
The ELAIS N2 field has also been observed:

(i) At radio (1.4 GHz) frequencies using the VLA, reaching a $3\sigma$
threshold  for source  detection $F(1.4~{\rm  GHz})\simeq  27.6\rm \mu
Jy$, with resolution  1.4 arcsec (see
Ivison et al. 2002).

(ii) In X-rays using the Chandra  satellite, for 75 ks in August 2000,
reaching source detection limits $F(0.5$--$8.0~\rm keV)\simeq 1.5\times
10^{-15}$  ergs $\rm s^{-1}cm^{-2}$  with sub-arcsec  resolution (
Manners et al. 2002).

    (iii) At $850\rm \mu m$ using SCUBA, reaching a $3.5\sigma$ source
  detection limit of $F(850\rm \mu m)\simeq 8$ mJy (Scott et al. 2002).

Here we discuss  these observations
 for the $K\leq 19.5$ UFTI  EROs only.  Seven of these 31 galaxies are
  detected at  1.4 GHz sources,
only one as an  X-ray source, and none in the sub-mm. Table 4 gives
 the fluxes of the detections.
\begin{table}
\caption{Radio and/or X-ray fluxes of the seven  $K\leq
19.5$ UFTI subsample EROs detected in the VLA and/or Chandra surveys.}
\begin{tabular}{lccc}
\hline
ERO & \multispan{2} \hfil $F(1.4~{\rm GHz})$ ($\rm \mu Jy$) \hfil &
 $F(0.5$--$8.0~\rm keV)$ \\
\smallskip
   & peak & integral & $10^{-15}$ ergs $\rm s^{-1}cm^{-2}$ \\
r256 & $~32.4\pm 9.5$ & $~25.1\pm 13.9$ & - \\
r518 & $~27.7\pm 9.5$ & $~30.0\pm 17.3$ & - \\
r552 & $~39.3\pm 9.5$ & $~21.8\pm 11.3$ & - \\
r608 & $4477.\pm 9.5$ & $5074.\pm 17.8$ & $3.060\pm 0.891$ \\
r642 & $~58.3\pm 9.5$ & $~46.7\pm 14.2$ & - \\
r660 & $~29.1\pm 9.3$ & $~39.3\pm 19.8$ & - \\
r952 & $~48.0\pm 9.5$ & $~43.4 \pm 15.4$ & - \\
\hline
\end{tabular}
\end{table}
In r256, r518, r552, r642 and r952, the radio emission is
 relatively weak  and consistent with  a point source, and  the host galaxies
 appear to be  regular ellipticals (the outer regions of r518
 are mildly disturbed). The radio  emission is
 stronger  than  would  be  expected  for  typical  $L^*$  ellipticals
 ($<10\rm \mu Jy$  at all $z>1$) and seems more  likely to be produced
 by  weak AGN, below  the Chandra  threshold, rather than 
 powerful starbursts.

The  galaxy r660, similar to these five in radio flux,  differs 
in that (i) the radio  emission is elongated, with major axis ${\rm
FWHM}\sim 2.3$ arscec on a  position angle $27\pm 18$ degrees, and (
 Figure  18)  aligned with the $K$-band long axis, (ii) the
galaxy is a large low-surface brightness disk. This suggests that  the
radio  emission   is produced by an extended
  dust-reddened  starburst. With  the Condon  (1992) radio  SED, the
$39.3\rm  \mu Jy$ flux corresponds  to a  rest-frame luminosity
($\nu L_{\nu}$) at $z=1(2)$ of $L_{1.4}=10^{39.66}(10^{40.34})h_{50}^{-2}$
 ergs $\rm s^{-1}$. If $L_{1.4}/{\rm  SFR}$ is in the range bracketed by
the Carilli (2000)  and Condon (1992) relations, as in Roche et
al. (2002),
the  total ${\rm SFR}=
99$--$233(474$--$1115)   h_{50}^{-2}\rm  ~M_{\odot}yr^{-1}$  at
$z=1(2)$, and r660 is a very
powerful starburst galaxy, probably similar to local ULIRGs.
\begin{figure}
\psfig{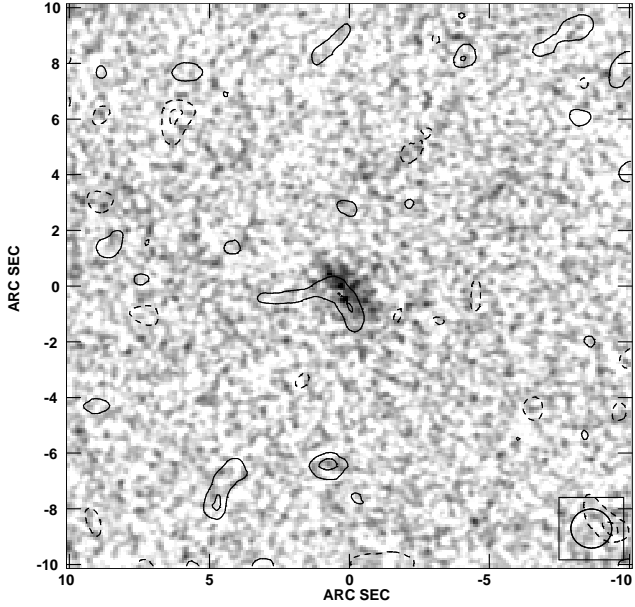}
\caption{Radio contour  map centred on  r660, generated from  VLA data
using the AIPS  package, superimposed on the UFTI  $K$-band image. The
radio  coutours are  at 2  and  3$\sigma$ where  $\sigma =9.2\rm  \mu
Jy$/beam (dotted contours are negative).}
\end{figure}

The non-detection of any of these 31 EROs in  the SCUBA survey
 sets an upper  limit on their SFR. Scott et  al. (2002)
estimate  the detection  limit of  $F(850\rm  \mu m)\simeq  8 mJy$  to
correspond to  a SFR  $\sim 1800  h_{50}^{-2} \rm ~M_{\odot}yr^{-1}$,
at all $1<z<10$, for the Salpeter IMF, which for the IMF of our models
becomes $\sim 1200 h_{50}^{-2} \rm ~M_{\odot}yr^{-1}$.
Most  of these  bright EROs, including the double nucleus and
disturbed galaxies, are  not detected  in the  VLA data either,
which sets stricter upper limits on SFRs, $\simeq 70$--$164
(335$--$785)h_{50}^{-2}\rm~M_{\odot}yr^{-1}$ at  $z=1(2)$. As most of
the  brightest EROs will be closer to $z\sim 1$, this implies that the
average SFR of dsfEROs at these redshifts is less than $\sim 200
h_{50}^{-2}\rm ~M_{\odot}yr^{-1}$. Nevertheless,
 a significant
minority -- e.g. r660, and the numerous examples of 
fainter EROs associated with SCUBA sources (Smail et
 al. 1999; Dey et al. 1999; Ivison et al. 2001; Lutz et al. 2001) -- 
are real ULIRGs.

By far the strongest radio source in this
sample, at  5mJy, is r608, the X-ray source and possible cluster  central
galaxy.
The  X-ray  and  VLA data (Figure 19) show most of the emission
from a point-source  concentric with  the $K$-band
profile. The radio flux is far too strong to be attributed to
star-formation  (the non-detection  with SCUBA limits the starburst
contribution to   $\leq 1$  per cent of this). If  r623 is at $z\simeq  1$, the
rest-frame 
radio  luminosity,   $L_{1.4}\simeq  10^{41.75}$  ergs  $\rm
s^{-1}$. This is
  3 orders of magnitude above  that of normal spirals
but 3 orders below the  most powerful radio galaxies. If we assume passive
optical evolution  and $(1+z)^3$ radio evolution for this galaxy
then at $z=0$ it
would  have  $M_R=-23.24$  and  $L_{1.4}\simeq 10^{40.85}$  ergs  $\rm
s^{-1}$, or $P_{1.4}=10^{24.70}$ W  $\rm Hz^{-1}$, and its position on
the radio-optical  luminosity plane (Ledlow and Owen  1996) would then 
be typical of local  FRI radio  galaxies. 
 
\begin{figure}
\psfig{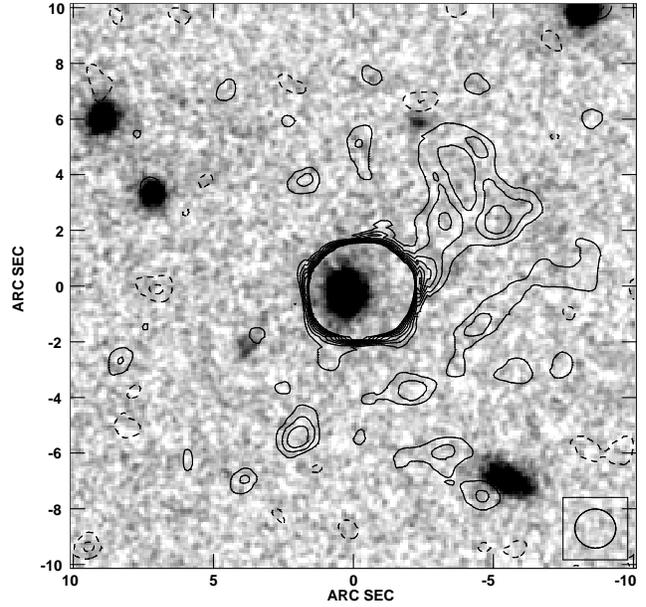}
\caption{Radio contour map centred on r608, 
superimposed on the UFTI $K$-band image. The radio contours are at 
  2,3,4...10$\sigma$ where $\sigma =9.2\rm \mu Jy$/beam.}
\end{figure}
The X-ray flux, $F(0.5$--$8.0~\rm  keV)= 3.06\times 10^{-15}$ ergs $\rm
s^{-1}cm^{-2}$,  corresponds  at   $z\simeq 1$,  assuming  a  $f_{\nu}\propto
\nu^{-1}$ spectrum, to $L_X(0.5$--$8.0{\rm keV})=10^{43.50}$ ergs $\rm
s^{-1}$, which  could also  be expressed in  the $\nu L_{\nu}$  form as
$L_{1\rm KeV}=10^{43.27}$ ergs $\rm s^{-1}$.  From this, the radio to X-ray
spectral index, $\alpha_{RX}\simeq{41.75-43.27+8.24\over 8.24}=0.815$,
which is  also consistent with an  FRI radio galaxy  (e.g.  Capetti et
al. 2002).
\section{Summary and Discussion}
\subsection{Summary}
\noindent (i) The number counts of EROs are significantly lower than
expected if all local E/S0s evolved as in a PLE
model, but are much higher than the predictions of  
 CDM-based  hierarchical merging models (Cimatti 2002b). We find
(in agreement with Firth et al. 2002) that the ERO counts are
consistent with completely non-evolving model for E/S0s.
 We can also fit the ERO counts with a
 physically plausible evolving model,`M-DE', in which 
both the characteristic mass
and the comoving number density of passively evolving galaxies
decrease with redshift.  Physically
this evolution could be accounted for by the continual formation of new 
passive  ellipticals from the
mergers and interactions of star-forming spirals.

Our best-fitting `M-DE' model 
 incorporates an observationally-estimated  merger rate 
(Patton et al. 2001), with the negative density
evolution parameterized by
$R_{\phi}=-0.46$,  meaning that 
only $\sim 35$ per cent of the present-day comoving number density of 
 E/S0 galaxies formed at
$z>3$. This fraction is higher than the $\sim
15$ per cent estimated by  Stiavalli and Treu (2001), because we have 
(a) included all types of ERO as possible E/S0 progenitors (b)
included a reduction in mean mass galaxy and hence $L^*$
at high redshift, due to merging
 thus allowing a
higher number density ($\Phi^*$)of passive galaxies.
\vskip 0.3cm
\noindent (ii) We examined the morphology of a bright ($K<19.5$)
subsample of 32 EROs on the UFTI mosaic. One appeared to be a Galactic
star, the other 31
a $\sim$3:2 mixture of elliptical and disk profile types. About 1/4 
showed some  evidence of asymmetry or disturbance, e.g  from recent
interactions, but only two were obvious mergers with double nuclei.
This mixture of morphologies is consistent  
 with the findings of
 Moriondo et al. (2001) and Stiavelli and Treu
(2001) for similar samples of EROs.

By fitting radial profiles, we estimated  seeing-corrected
half-light
radii. The $r_{hl}$ of the bulge-type EROs were
consistent with passively evolving E/S0 galaxies in the redshift range
($0.9<z<2.0$) predicted
by the M-DE model. This is in agreement with
 Moriondo et al. (2000) who found the surface brightness--size
relation of six bulge-type EROs at known redshifts of 
 $z\sim 1.3$ to be consistent 
 with  passively evolved ($\Delta(B)=-1.4$ mag)  ellipticals.

Although both the M-DE and non-evolving models fit the ERO
counts, these models give different angular size distributions, and
our observed ERO radii are much more consistent with the smaller radii
predicted by M-DE. This implies that, whatever the $L^*$ evolution,
 the surface brightness evolution of EROs is at least that predicted
for passive evolution, and if the $L^*$ evolution is much weaker this
must be the result of merging.
 The average $r_{hl}$ of the disk-profile EROs is even smaller than
the M-DE prediction and possible explanations include inside-outwards
disk formation and/or morphological evolution. 
\vskip 0.3cm
\noindent(iii) Radio observations with the VLA detect emission above the
$3\sigma$ limit of $F(1.4~{\rm  GHz})\simeq  27.6\rm \mu
Jy$ for 7 of the 31 ERO galaxies with  $K<19.5$ on the  UFTI field.
The strongest  of these sources, at 5mJy, would be a radio galaxy,
perhaps like local FRIs 
if these evolve as strongly as $L_{rad}\propto(1+z)^3$.
The host galaxy is also a Chandra X-ray source and 
may be the  central giant elliptical  of a
cluster of EROs. Of the other, much fainter (30--$60\rm \mu Jy$)
detections, five are point sources within
apparently regular galaxies, and may be weak AGN. The seventh, the
galaxy r660, shows elongated radio emission aligned with the
$K$-band image. The galaxy is a large disk of lower surface brightness
than the other bright EROs. These properties suggest it is a powerful
but very dust-reddened starburst, with   a                 SFR
100--1000 $h_{50}^{-2}\rm ~M_{\odot}yr^{-1}$. 

The non-detection of the other EROs, which include double-nucleus and
disturbed objects, implies their SFRs are lower.
This was interpreted
as indicating the mean SFR of the dsfEROs to be
$< 200h_{50}^{-2}\rm ~M_{\odot}yr^{-1}$. This is probably
 consistent with the Cimatti et al. (2002a) estimate of 
$\sim 100
h_{50}^{-2}\rm ~M_{\odot}yr^{-1}$, based on 
the mean $\rm[OII]3727$ and UV fluxes of spectroscopically observed
dsfEROs (see 8.2 below), after correcting for $E(B-V)=0.5$ mag extinction. 
     \vskip 0.3cm
\noindent (iv) We investigated the angular
 correlation function $\omega(\theta)$ of 
EROs on the UFTI and Ingrid fields. Positive clustering is detected at
$>2\sigma$ for the EROs to $K=19.5$-20.0, with an estimated 
$\omega(\theta)$ amplitude about an order of
magnitude higher 
than  that of all galaxies to
the same $K$ mag limits. Our ERO clustering results are consistent
with those of 
 Daddi et al. (2000) and Firth et al. (2002), and using a combination
of these, we estimate that the brighter ($K\leq 20.0$) EROs
have a comoving correlation radius $r_0\simeq 10$--13$  ~h_{100}^{-1}$
Mpc, depending on the strength of $L^*$ evolution. 
Statistical
uncertainties remain large and it there is clearly a need for further
ERO clustering analyses with both larger and deeper samples.

\subsection{Discussion}
\subsubsection{Star-forming EROs}
Cimatti et al. (2002a), using VLT spectroscopy, classified
the spectra of  about two-thirds of a sample of bright ($K\leq 19.2$) 
EROs, and found approximately half ($50\pm 17$ per cent) to be 
dusty star-forming
 and half to be old ($>3$ Gyr) passive galaxies.
 The averaged spectrum of the dsfEROs
 showed [OII]3727 emission and Balmer absorption lines and
closely resembled an e(a) type (Poggianti and Wu 2000) `very luminous infra-red galaxy' with
stellar reddening $E(B-V)\sim 0.5$.

The VLA data indicate that only a small fraction
of EROs can be ULIRGs, and furthermore
the proportion of EROs which are ongoing  mergers or
very  disturbed is considerably lower than
 50 per cent, so 
some dsfEROs must have more regular bulge or intermediate
 profiles. This suggests that many
of the dsfEROs are
in late post-merger  stages, with low and declining 
SFR and increasingly regular
morphology. These galaxies may subsequently   
become  E/S0s and contribute to the   
increase with time in the comoving number density of passive galaxies.
 The dsfEROs and pEROs  would then be 
separate stages of an evolutionary sequence. At  
$z\sim 1$--2 we
see the second (pERO) stage for the earliest formed ellipticals,
contemporaneous with the dsfERO stage of those forming later, while
deeper ERO surveys would reach the  
dusty starburst phase of the first massive 
ellipticals, at $z\geq 3$ (i.e. the SCUBA sources).

On the basis of the  Cimatti et
al. (2002a) spectroscopy,  
dsfEROs would  make up $\sim(0.5\pm 0.17)\times 14 = 7\pm 2$
per cent of all $K\leq 20$ galaxies and to this limit have a surface
density $0.69\pm 0.23$ $\rm arcmin^{-2}$. These high numbers imply the
dsfERO phase is prolonged. For  an order-of-magnitude estimate of the
total
 dsfERO lifetime, $t_{dsf}$, we  assume that (i) each
present-day E/S0 galaxy  is associated with an average of one dsfERO `event' at
higher redshift, (ii) that the $K$-band magnitudes of dsfEROs can
be approximated by the passive model with merging (as
they seem to be similar in surface brightness to the pEROs).

If $t_{dsf}$ were to last for the entire `ERO epoch', $z>0.93$, the dsfERO
count could then be represented by the $z>0.93$ E/S0s in a model with 
merging mass evolution ($R_m=0.3$)  but a comoving number density
remaining at the present0day value 
($R_{\phi}=0$). This would be intermediate between the merging and M-DE
models on Figure 6 and to  $K=20$ gives 2.9 $\rm arcmin^{-2}$. The
observed dsfERO count is lower, implying a shorter $t_{dsf}$. As most
 $K\leq 20$ EROs will be at $0.93<z<2$, or lookback times in the period
$(10.4$--$14.4)h_{50}^{-1}$ Gyr, hence $t_{dsf}\sim
{0.69\over 2.3}\times (14.4-10.4)~h_{50}^{-1} = 1.2~h_{50}^{-1}$ Gyr.
\subsubsection{Clustering and evolution of the EROs}
 We estimate the intrinsic  clustering of bright ($K\leq 20$) EROs as 
 $r_0\simeq 10$--$13~h^{-1}_{100}$
Mpc comoving, depending on the strength of $L^*$ evolution. This is
even  stronger than the clustering of present-day giant ellipticals,
 $r_0\simeq 8~h^{-1}_{100}$, and  implies that
if the $z\sim 1$--2  EROs
evolve into the $z\sim 0$ E/S0s, 
 the increase in their comoving number density must involve
 the assimilation of less clustered types of galaxy 
into the class of passive galaxies, 
reducing the overall $r_0$.

It may be significant that Abell galaxy clusters, 
with $r_0\simeq 21 h^{-1}_{100}$
Mpc (Abadi, Lambas and
Muriel 1998), and possibly SCUBA sources (Almaini et al. 2002), may be
even more clustered than the $z\sim 1$--2 EROs. On the other hand, the 
Lyman break galaxies at $z\sim 3$ are on the
whole more moderately
 clustered, with a comoving 
 $r_0\simeq 2$--$5~
h_{100}^{-1}$ Mpc (Giavalisco and Dickinson 2001; Arnouts et
al. 2002), which suggests they are mostly the 
 progenitors of   
disk galaxies.

Daddi et al. (2002) found, on the basis of  a small sample of 
redshifts, some evidence  that the strong ERO clustering 
is associated with the
pERO types only. They estimate 
$r_0<2.5~h_{100}^{-1}$ Mpc for dsfEROs, and  
argue that this may be
 evidence against dsfEROs being E/S0 progenitors.

However, a large
difference in pERO and dsfERO clustering would  be expected 
if (i) the pEROs
form at very dense mass concentrations in the early Universe, 
the `seeds' of 
 Abell clusters and SCUBA sources (and so have   similar clustering),
and (ii) the  addition with time of
dsfEROs (formed in mergers of the relatively weakly clustered disk
galaxies)
to the ERO class produces a continual dilution of the
clustering -- until, by $z\sim 0$,   the passive galaxy  $r_0$ is as
observed for local  E/S0s.

Secondly, IRAS ($60\rm \mu m$) selected galaxies,
 which may be considered as low redshift
counterparts of the dsfEROs, are   weakly
clustered ($r_0\simeq 4~h_{100}^{-1}$ Mpc), but nevertheless 
have many properties suggesting  they are destined to evolve into   
 E/S0s (Genzel et al. 2001).

Thirdly, many low-redshift E/S0 are observed to have have `boxy' or
`disky' isophotes, and this may be `fossil' evidence that  
they evolved from  both types of ERO.
 The 'boxy' ellipticals tend to be more massive and
more confined to the centres of clusters, with slightly older stellar
populations.
 The 'disky' galaxies tend to have
 higher rotation velocities ($v_{rot}$) but lower internal velocity
 dispersions ($\sigma_i$). 
 These differences
may result from the `disky' ellipticals having formed  at a relatively late
  epoch, when large
spirals 
were already present -- perhaps (at one extreme) from the merger of a
single pair of spirals, which 
 is likely impart a substantial angular momentum to the
post-merger. At the other extreme, the oldest
ellipticals   
might be formed  through multiple mergers of very many small
disks in the core of a proto-cluster, 
giving a  summed 
angular momentum vector much closer to zero, and `boxy' isophotes  
This latter process may have been observed in
 deep $K$ images of high-$z$ radio galaxies, which show 
massive ellipticals at $z<3$ but  a large number of clustered
sub-components at $z>3$ (van Breugel et al. 1999).
 
At $z\sim 1$--2, the progenitors of today's `boxy' and `disky' E/S0s would
then be, respectively, the pEROs and dsfEROs.
\subsubsection{Future plans}
We hope to investigate these scenarios further through 

(i) Analysing the clustering and morphologies of 
of larger samples
 of EROs, with multi-colour
imaging (e.g. $R$, $I$, $J$, $H$ and $K$) to enable separation of the 
dsfEROs and pEROs, to fainter ($K\geq 22$) limits. We now have some
data for this. 
It may prove especially useful to compare the clustering properties of EROs and
 SCUBA sources. 

(ii) High-resolution 
spectroscopy of a diverse sample of EROs, with the aim of not only
determining the $N(z)$, but also
of estimating properties such as current SFR, internal dust extinction, metallicity and
kinematics, and applying new techniques of  
  age-dating the stellar populations. From  the velocities  $\sigma_i$ and 
$v_{rot}$, measurable from ground-based
spectroscopy, on a 
$v_{rot}$--$\sigma_i$ plot (Genzel et al. 2001),
 the ERO progenitors of `disky'
and `boxy' ellipticals might  be identified. 
\subsection*{Acknowledgements}
This paper is based on observations with the WHT and UKIRT.
The William Herschel Telescope is operated on the island of La 
Palma by the Isaac Newton Group in the Observatorio del Roque de los 
Muchachos of the Instituto de Astrofisica de Canarias. The United
Kingdom Infrared Telescope is operated by the Joint Astronomy Centre
on behalf of the UK Particle Physics and Astronomy Research Council. 
NR acknowledges the support of a PPARC Research Associateship. 
OA  acknowledges the support of a Royal Society  Research Fellowship.
JSD acknowledges the 
support of a PPARC Senior Fellowship. 
CJW and RJI thank PPARC for support. 
  
\end{document}